# Nonlocal Optomechanics: Hybrid Anapole Opens a New Route to Optical Tweezing

Susanna R. Rozental[1], Denis A. Kislov[1,3], Ilia M. Fradkin[1,3], Nikita S. Babich[2], Vasiliy Fedotov[3], Sergey Novikov[1], Vjaceslavs Bobrovs[4], Shangran Xie[5], Oleg Minin[6], Igor Minin[6], Lei Gao[7], Yu-Ling Wu[8], Lei Gong[8], Alexey Bolshakov[1], Alexey Arsenin[9], Alexander S. Shalin[1,3,7,10]

[1] Moscow Center for Advanced Studies, Kulakova str. 20, Moscow 123592, Russia
[2] ITMO University, Kronverkskii 49, St. Petersburg 197198, Russia
[3] Skolkovo Institute of Science and Technology, Bolshoi Boulevard 30, 121205 Moscow, Russia
[4] Riga Technical University, Institute of Photonics, Electronics and Telecommunications, 1048 Azenes street 12, Riga, LV 1658
[5] Beijing Institute of Technology, Beijing 100081, China
[6] Tomsk Polytechnic University, Lenina 36, Tomsk, 634050, Russia
[7] School of Optical and Electronic Information, Suzhou City University, Suzhou 215104, China
[8] Department of Optics and Optical Engineering, University of Science and Technology of China, Hefei 230026, China
[9] Emerging Technologies Research Center, XPANCEO, Internet City, Emmay Tower, Dubai, United Arab Emirates
[10] Faculty of Physics, M. V. Lomonosov Moscow State University, Moscow 119991, Russia

Optical tweezers confine a particle in an intensity-defined potential well by engaging its local multipoles. In this picture, eliminating far-field scattering from the particle should cancel the optical force, as the multipole moments underpinning the conventional optomechanical response vanish. We show that certain resonant states, such as, e.g., the hybrid anapole state, enable qualitatively different optical manipulation, nonlocal by nature, where the optical force exhibits nontrivial spatial variations absent in conventional tweezing, establishing a new framework for manipulating resonant nanoparticles.

*Introduction.* – Optomechanical manipulation, pioneered by A. Ashkin [1–3], has evolved into a versatile and powerful tool for trapping and guiding micro- and nanoscale objects across biophysics [4–7], quantum physics [8–10] and lab-on-a-chip-microfluidics [11–13]. Conventional optical tweezers theory for small nanoparticles treats optomechanical response as local [14–16]: dipolar/multipolar moments induced in a particle are determined by the incident field and its first derivative at the particle's center of mass. Therefore, an effective optical potential (proportional to the incident light intensity (Fig. 1a)) governs conservative transverse optical gradient force. Consequently, in most cases (besides specially designed platforms [17–21]) a Gaussian beam generates a simple potential well, which defines the standard trapping configuration and ensures a stable equilibrium position [22] (Fig. 1b).

However, this potential-based description is not universal. The limits of conventional optomechanics are achieved by suppressing far-field scattering, thereby eliminating the dominant radiative channels that normally govern the optical force. This can be achieved with the hybrid anapole state, when the Cartesian primitive multipole moments induced in the particle are almost completely canceled by their corresponding toroidal counterparts [23–25]. At first glance, this multipole compensation suggests that the trapping optical force should also vanish.

Here we demonstrate exactly the opposite: suppressing far-field scattering opens a fundamentally new nonlocal optomechanical regime, in which the optical force not only persists but also exhibits nontrivial spatial variation (Fig. 1c,d) unattainable in conventional optical traps. We elucidate the origin of this phenomenon by introducing higher-order polarizability tensors associated with optical field gradients up to fourth order. Our analysis reveals the dominant role of this high-order contributions to optical forces and predicts qualitatively new optomechanical effects, including off-axis trapping, two-point equilibria, and asymmetric trapping featuring a spatially extended zero-force plateau, thereby enabling new strategies for optical tweezing and nanoparticle sorting in microfluidic and lab-on-a-chip systems.

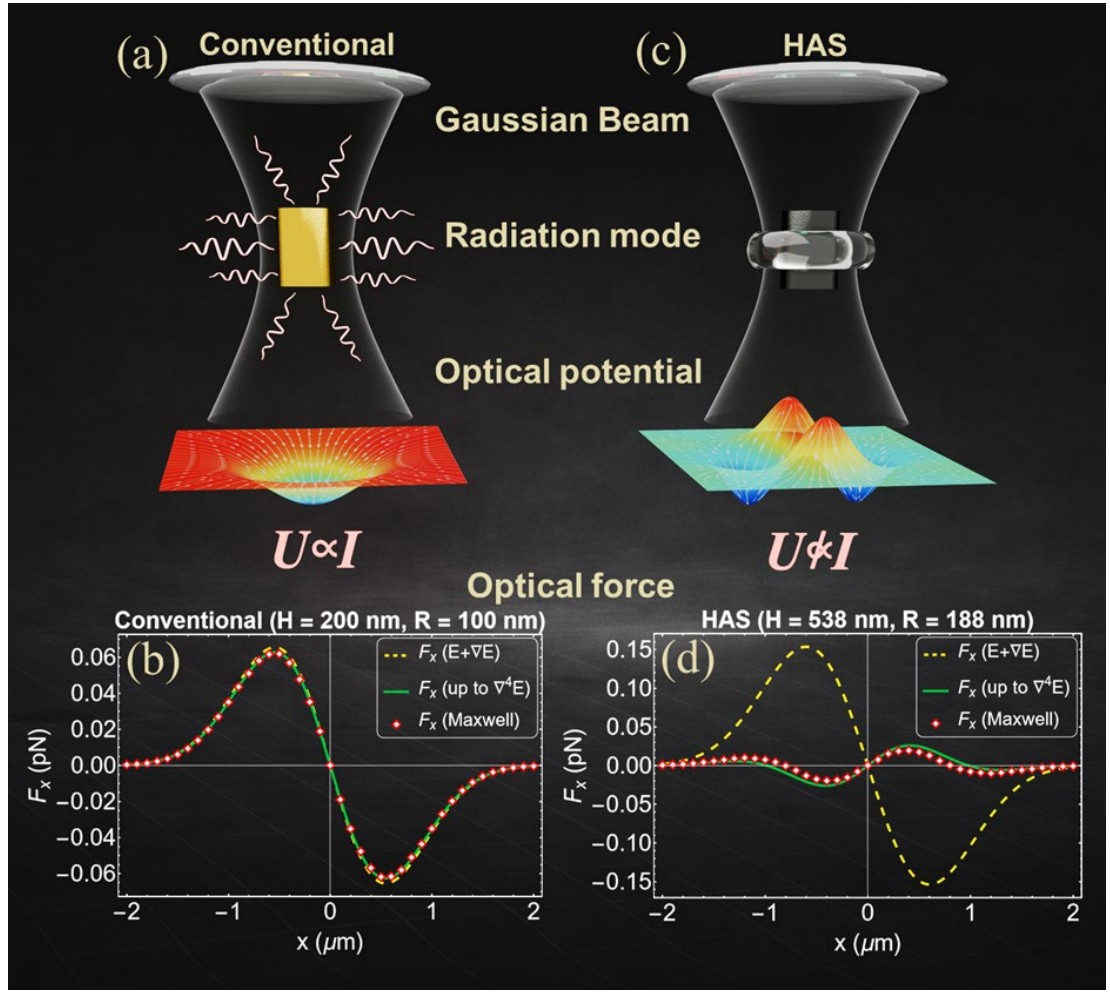

Fig. 1. From local to nonlocal optomechanics: (a) Classical optical tweezers: the interaction of a Gaussian beam with a particle exhibiting a local response creates an effective potential well with $U \propto I$, into which the particle is drawn and trapped. (b) Transverse optical force profile in conventional local models, which implicitly assume that the particle responds only to the field at its center. (c) In the hybrid anapole state, far-field scattering is strongly reduced, effectively disabling radiative channels and allowing nonlocal contributions driven by incident field gradients to dominate the optical force. (d) Transverse force profile in the nonlocal regime: higher-order spatial derivatives of the field (up to fourth order) reshape the force landscape, giving rise to nontrivial force distributions and new types of stable equilibria. Panels (b) and (d) compare three levels of theory: the yellow dashed curves show the local model obtained from Eqs. (1) and (2); the green solid curves show our nonlocal theory based on Eq. (3), including higher-order gradient terms; and the red markers give the full-wave reference result from numerical integration of the Maxwell stress tensor.

*Hybrid Anapole State.* – While the conventional electrodynamic anapole corresponds to a scattering minimum in a single multipole (typically dipole) channel [26,27], the contributions from other multipoles typically prevent the complete suppression of far-field radiation. By engineering the particle geometry, however, the scattering minima for all multipolar channels can be spectrally overlapped, achieving an almost vanishing far-field response [24,28]. This state is known as the hybrid anapole state (HAS) (for details, see Supplementary, Sec. 2). In what follows, we examine the peculiar optomechanics of a particle supporting such a state within a standard optical trap.

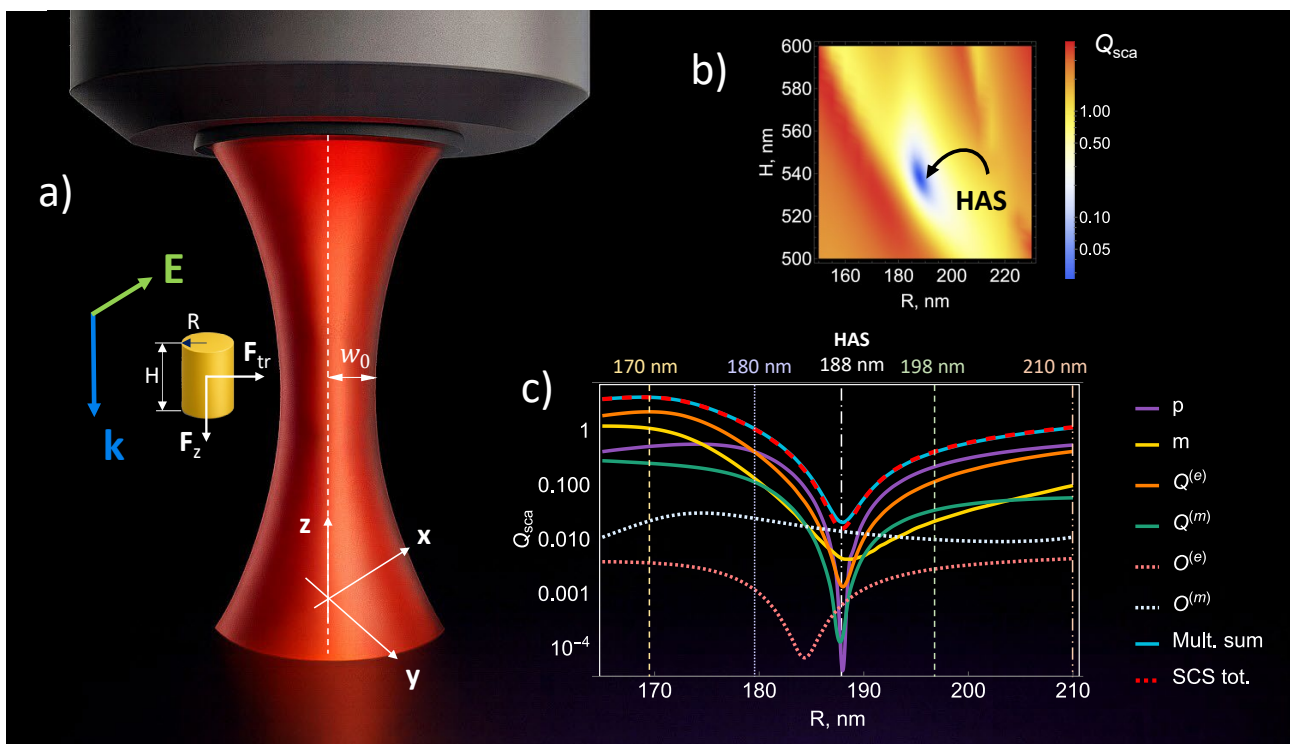

Fig. 2. a) A schematic of the investigated setup: cylindrical nanoparticle placed in the focal plane of the laser beam and displaced laterally from its axis. Therefore, both optical pressure $F_z$ and a transverse optical force $F_{tr}$ govern its dynamics; b) The scattering cross section of the nanocylinder has been normalized to its geometric cross section ($Q_{sca} = \frac{\sigma_{sca}}{\pi R^2}$) and shown as a function of height H and radius R; c) The spherical multipole decomposition for the nanoparticle with fixed height H =

538 nm as a function of its radius (p – electric dipole, m – magnetic dipole, $Q^{(e)}$ – electric quadrupole, $Q^{(m)}$ – magnetic quadrupole, $O^{(e)}$ – electric octupole, $O^{(m)}$ – magnetic octupole). The vertical lines correspond to different radii of the nanoparticles, which exhibit different optomechanical patterns, including the hybrid anapole state (R = 188 nm).

We consider a silicon nanocylinder in air positioned in the focal plane of a linearly polarized Gaussian beam (Fig. 2(a)). The wavelength is fixed at $\lambda$ = 1064 nm (YaG:Nd$^{3+}$), while the nanocylinder's radius R and height H serve as tuning parameters. The beam waist is equal to the wavelength $w_0 = \lambda$; light propagates along the z-axis and is polarized along the x-axis. The field amplitude at the beam center is $E_0 = 10^6$ V/m, which is typical for conventional optical tweezers.

Fig. 2b reveals a region in (H,R) space, where the total scattering is almost completely suppressed with the absolute minimum occurring at H=538 nm and R=188 nm. Fig. 2c shows the normalized scattering cross section, $Q_{sca}$, and its multipole decomposition up to the octupole near this minimum as functions of the radius [29,30]. As *R* approaches 188 nm, all dipole and quadrupole contributions are seen to decrease substantially (see Supplementary, Sec. 3.4), making the total scattering weaker by nearly 200 times and, as the result, leading to the hybrid anapole state. Moreover, Fig. 2c shows an excellent agreement between the scattering cross sections calculated by integrating the Poynting vector over a surface enclosing the nanocylinder and by summing all multipole contributions. Remarkably, in the HAS spectral band the magnetic octupole, which is usually neglected, provides the dominant scattering contribution. Nevertheless, as we show below, it doesn't affect the overall optical force landscape.

*Generalized Nonlocal Approach* – The optical force acting on a particle is determined by the multipoles supported by the latter and by the interaction of induced displacement currents with the driving electromagnetic field. With this in mind, the time-averaged optical force can be written as follows [31]:

$$F_i = \frac{1}{2}\operatorname{Re}[p_j \nabla_i E_j^*] + \frac{1}{12}\operatorname{Re}[Q_{jk}^e \nabla_i \nabla_k E_j^*] + \frac{1}{12}\operatorname{Re}[O_{jkl}^e \nabla_i \nabla_l \nabla_k E_j^*] +$$

$$+ \frac{1}{2}\operatorname{Re}[m_j \nabla_i B_j^*] + \frac{1}{12}\operatorname{Re}[Q_{jk}^m \nabla_i \nabla_k B_j^*] + \frac{1}{12}\operatorname{Re}[O_{jkl}^m \nabla_i \nabla_l \nabla_k B_j^*] -$$

$$- \frac{k^4}{12\pi\varepsilon_0 c}\operatorname{Re}[\varepsilon_{ijk} p_j m_k^*] - \frac{k^6}{2160\pi\varepsilon_0 c^2}\operatorname{Re}[\varepsilon_{ijk} Q_{lj}^e (Q_{lk}^m)^*] - \frac{k^8}{7560\pi\varepsilon_0 c}\operatorname{Re}\left[\varepsilon_{ijk} O_{lnj}^e (O_{lnk}^m)^*\right] -$$

$$- \frac{k^5}{120\pi\varepsilon_0}\operatorname{Im}[Q_{ij}^e p_j^*] - \frac{k^7}{1890\pi\varepsilon_0}\operatorname{Im}[O_{ijk}^e (Q_{jk}^e)^*] -$$

$$- \frac{k^5}{120\pi\varepsilon_0 c^2}\operatorname{Im}[Q_{ij}^m m_j^*] - \frac{k^7}{1890\pi\varepsilon_0 c}\operatorname{Im}[O_{ijk}^m (Q_{jk}^m)^*],$$

(1)

where $\varepsilon_{ijk}$ is Levi-Civita symbol and indices *i, j, k* and *l* correspond to coordinates $\{x,y,z\}$, *E* is the incident electric field, and *B* is the incident magnetic field.

The first six terms in Eq. (1) describe the so-called interception forces arising from the interaction of each multipole with the electromagnetic field. The remaining terms represent recoil forces originating from the interaction between pairs of different multipoles.

The calculation of the force using Eq. (1) comes down to determining the multipole moments induced in a particle. In conventional scattering regimes the linear response is considering [15] and these moments are given by the incident field and its first derivatives at the particle's center multiplied by the corresponding polarizability tensors $\alpha_p, \alpha_m, \alpha_{Q^e}, \alpha_{Q^m}$ [16,32]:

$$\mathbf{p} = \alpha_p \mathbf{E}_{inc}; \quad \mathbf{m} = \alpha_m \mathbf{H}_{inc}$$
$$\mathbf{Q}^e = \alpha_{Q^e} \frac{\nabla \mathbf{E}_{inc} + \mathbf{E}_{inc}\nabla}{2}; \quad \mathbf{Q}^m = \alpha_{Q^m} \frac{\nabla \mathbf{H}_{inc} + \mathbf{H}_{inc}\nabla}{2},$$

(2)

For the hybrid anapole state this local theory yields the classical trapping force profile (Fig. 1d, yellow dashed curve), which deviates dramatically from the outcome of a full-wave simulation (Fig. 1d, red markers). The discrepancy extends to nearly all terms in Eq. (1), with the dipole and quadrupole forces also exhibiting nonclassical behavior (See Supplementary, Sections 3.5). Therefore, the conventional local description has a clear limit of validity and fails in the HAS regime.

Here we introduce a more general framework, which does not rely on the locality assumption, remaining valid even in special scattering regimes and revealing unusual optomechanical dynamics. To this end, we expand the incident field in a Taylor series and employed widely used scattered field formulation [33,34]. Thus, we obtained a hierarchy of generalized polarizability tensors, which link induced multipoles to higher-order field gradients [35–37]. The general form of the series corresponding to a multipole, M, including derivatives up to the N-th order reads:

$$M_i[\mathbf{E}(\mathbf{r})] = M_i \left[ \sum_{n=0}^{N} \sum_{k_1,k_2,k_3} \frac{\delta_{n,k_1+k_2+k_3}}{k_1!k_2!k_3!} \frac{\partial^n \left( \sum_{\beta=x,y,z} E_\beta \mathbf{e}_\beta \right)}{(\partial^{k_1} x)(\partial^{k_2} y)(\partial^{k_3} z)} x^{k_1} y^{k_2} z^{k_3} \right] = $$

$$= \sum_{n=0}^{N} \sum_{k_1,k_2,k_3} \frac{\delta_{n,k_1+k_2+k_3}}{k_1!k_2!k_3!} \sum_{\beta=x,y,z} \alpha^M_{ik_1k_2k_3\beta} \frac{\partial^n E_\beta}{(\partial^{k_1} x)(\partial^{k_2} y)(\partial^{k_3} z)}$$

(3)

where $\mathbf{e}_\beta$ is a unit vector along the $\beta$-axis that corresponds to coordinates $\{x,y,z\}$, indices $i = (i_1...i_n)$ denote a set of multipole components ($i$=1 –dipole, $i$=2 – quadrupole, etc.), $\alpha^M_{ik_1k_2k_3\beta}$ is a component of a polarizability tensor corresponding to the response of an $i^{th}$ component of the multipole M to the gradients of a β-th component of an incident field.

To elucidate the importance of this generalized gradient-based description, we consider two silicon nanocylinders shown in Fig. 1: one – in the conventional scattering regime, where the local approach works well, and another – in the hybrid anapole state, where the local approach apparently breaks down. For each case, we compute the transverse-force profile $F_x(x)$ in three ways, using: (i) full-wave numerical calculations based on the Maxwell stress tensor (COMSOL Multiphysics), (ii) local approximation retaining only an incident field and its first gradient, and (iii) our generalized formalism including derivatives up to the fourth order. In cases (ii) and (iii) we compute the optical force by inserting Eq. (3) into Eq. (1).

The symmetry of the system automatically eliminates a subset of terms in the expansion. In particular, for the electric dipole, magnetic quadrupole, and electric octupole nonzero contributions arise from even-order derivatives (i.e. $\mathbf{E}$, $\nabla^2\mathbf{E}$, and $\nabla^4\mathbf{E}$), whereas the magnetic dipole, electric quadrupole, and magnetic octupole interact with odd-order derivatives only (i.e., $\nabla\mathbf{E}$ and $\nabla^3\mathbf{E}$).

For a conventionally scattering nanocylinder (Fig. 1b), the local approximation (yellow dashed curve) reproduces the classical trapping profile $F_x(x)$ and agrees well with the full-wave numerical simulation (Fig. 1b, curve with red markers). In this case, the force is indeed governed by the local multipolar response and lowest-order field derivatives. In stark contrast, Fig. 1d shows that the local model (yellow dashed curve) deviates strongly from numerical calculations in the case of HAS-resonant regime (curve with red markers). Introducing nonlocal terms up to the fourth order in the frame of our approach yields excellent agreement with the full-wave numerical solution (see Fig. 1d green solid curve).

It is important to clarify why the local approximation still predicts a nonzero optical force in the HAS regime, despite failing to reproduce the full-wave results. While the first four multipoles (both electric and magnetic dipoles and quadrupoles) are strongly suppressed, higher-order multipoles (most notably the octupoles) remain prominent. In this regime, their scattering contributions (especially of the magnetic octupole) become higher than those of the suppressed lower-order multipoles (see Fig. 2c) and, thus, ensure non-vanishing optical force (see Supplementary Information, Sec. 3.4).

*Results and Discussion.* – It was previously shown that eliminating far-field scattering drives high-index nanoparticles into a qualitatively new optomechanical regime. In this case the transverse force no longer follows the conventional U∝I relationship and, instead of a single radially symmetric trap, can produce optical traps with nontrivial spatial landscapes. Here we examine how changing the nanocylinder's radius reshapes the optical traps near the hybrid anapole state.

We focus on the transverse force components, as they govern trapping stability and in-plane motion; the longitudinal force $F_z$ (radiation pressure) is analyzed in Supplementary, Sec. 3.3. Because of the axial symmetry of a Gaussian beam, the transverse force vanishes on the optical axis. To fully trace the force landscapes, we vary the nanocylinder position over ±1.5 μm along the x- and y-directions and analyze the profiles of $F_x(x)$ and $F_y(y)$ for the fixed height H=538 nm and different radii.

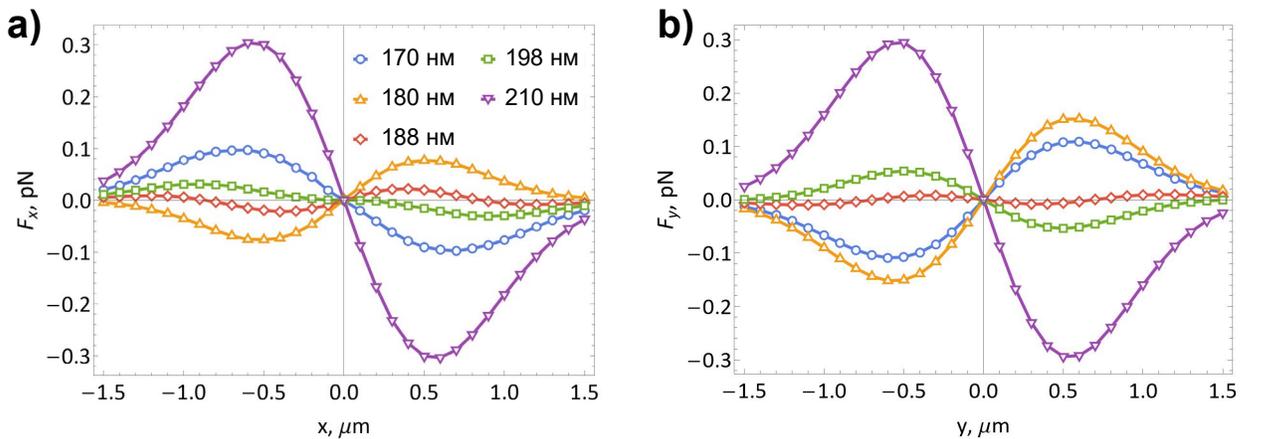

Fig. 3. Transverse optical forces. a) $F_x$ as a function of x-coordinate, b) $F_y$ as a function of y-coordinate. $F_x$ and $F_y$ are plotted for different radii: 170 nm (blue), 180 nm (yellow), 188 nm (red) (hybrid anapole state), 198 nm (green), 210 nm (purple).

The profiles of $F_x(x)$ and $F_y(y)$ are highly sensitive to the cylinder radius, especially close to the HAS (see Fig. 3). For the particle with R=210 nm, which is far from the HAS (purple curves in Fig. 3), both force components exhibit conventional trapping behavior: for small lateral displacements from the beam center the restoring force drives the particle back toward the optical axis. As the radius decreases, the force landscape is reshaped. At R=170 nm (blue curves in Fig. 3) we observe a mixed regime: trapping persists along one axis, whereas anti-trapping occurs along another axis, resulting from a saddle-like variation of the force in the plane. For R=180 nm (yellow curves in Fig. 3) both transverse components turn repulsive, corresponding to anisotropic anti-trapping. The transition from trapping to antitrapping is governed by the sign of the polarizability tensors of individual multipoles. This scenario is also achievable in other resonant particles, as shown in [17,18,21].

Near the HAS, at R=198 nm, trapping re-emerges, but exhibits a marked asymmetry: while a conventional trap with an equilibrium position at the beam center is formed along the y-axis (green curve in Fig. 3b), the force along the x-axis has a pronounced flattening near the beam center (green curve in Fig. 3a). Consequently, a spatially extended near-zero-force plateau develops around the optical axis, resulting in weak confinement of the particle. Finally, in the HAS case (R=188 nm, red curves in Fig. 3), both $F_x(x)$ and $F_y(y)$ change the sign at finite distances from the beam center on each side (on a submicrometer scale), giving rise to stable equilibrium positions away from the intensity maximum. This is an inherently unusual situation for conventional optomechanics driven by a Gaussian beam.

To connect the spatial variations of the force to particle motion, we model the nanocylinder dynamics in the focal plane by solving Langevin equations (see Supplementary Sec. 1.4 for details). In air, viscous damping and stochastic forces are weak, so the motion of the nanocylinder is chiefly governed by the optical force. For clarity, we plot the spatial maps of the quantity $U$ defined as a numerical integral of the transverse optical force $\mathbf{F}_{tr} = \mathbf{F}_x + \mathbf{F}_y$ over the focal plane (see Supplementary Sec. 1.5 for details):

$$U(x_0, y_0) = \int_{\infty}^{(x_0, y_0)} \mathbf{F}_{tr} \cdot d\mathbf{r} \qquad (4)$$

While $U$ does not represent the true potential energy due to possible nonconservative components of optical forces, it retains the dimension of energy and serves as a convenient indicator of the effective "motion landscape" and locations of stable attractors (see Fig. 4). For each radius, we initialize an ensemble of identical particles at random positions with zero initial velocity and identify the positions of attraction in the focal plane.

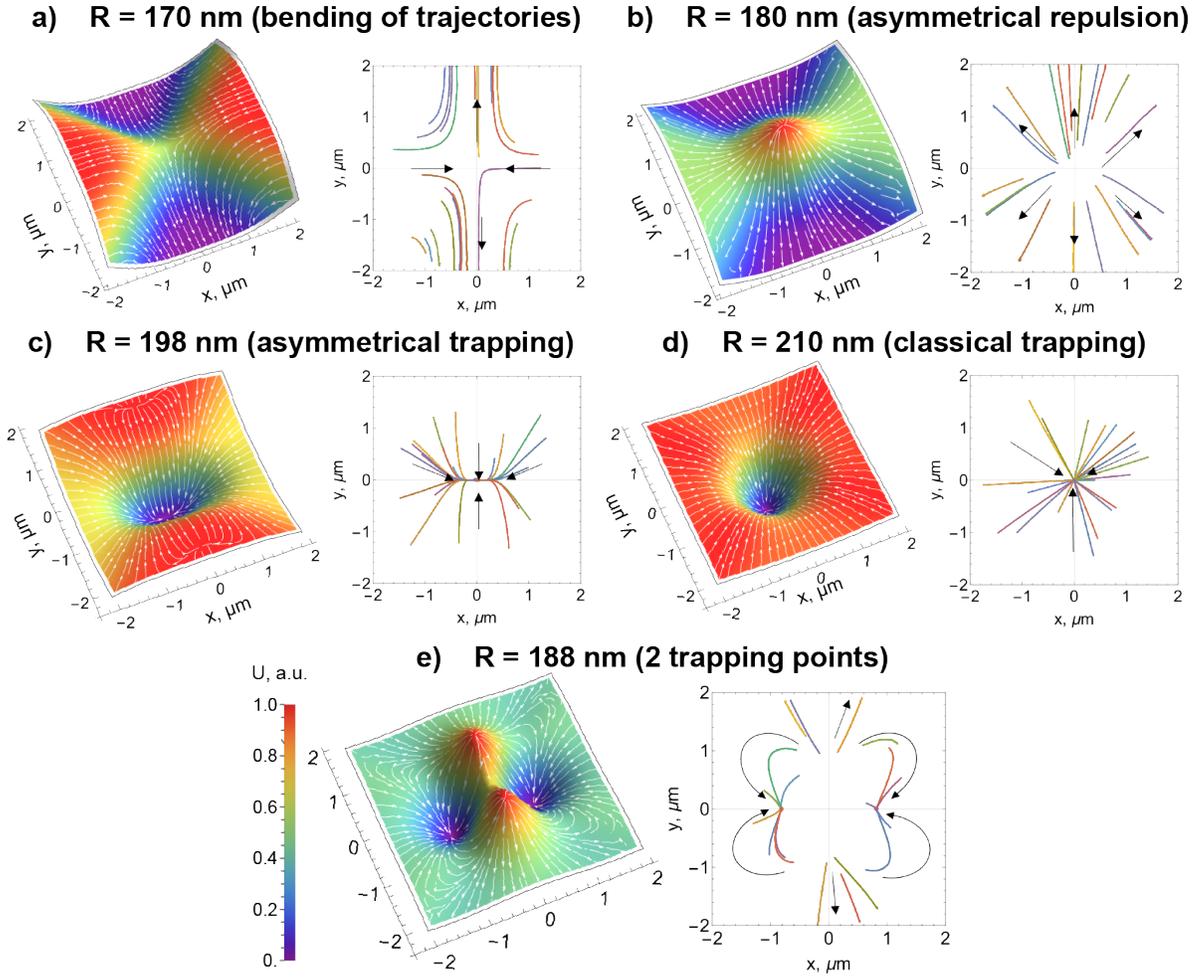

Fig. 4. Left panels (a–e): focal-plane distributions of the quantity U (an energy-dimensioned integral of the transverse force, according to Eq. (4)), serving as an effective particle "motion landscape". The maps are normalized to their maxima, with the minima set to zero. The right panels a)-e) show the trajectories of 20 nanocylinders of the same radius (for each panel) in the corresponding force fields. All the nanocylinders are suspended in air with randomly chosen starting positions and zero initial speed. The particle's movement time is 20 ms, the calculation step is 0.02 ms

The energy profiles and the corresponding trajectories of the particles shown in Fig. 4 confirm the existence of 5 different optomechanical effects near the HAS: saddle-shaped bending of trajectories (a), asymmetric repulsion (b), asymmetric trapping with zero-force plateau (c), classical trapping (d), and trapping with two equilibrium positions (e). It is important to emphasize that these optomechanical effects are not merely an exotic extension of conventional trapping, nor do they arise from fine tuning of the parameters within the same physical framework. Instead, they reflect a qualitatively different regime of optical-force formation. In ordinary scattering regimes the transverse-force spatial profile is governed by a local multipolar response, leading to a single centrosymmetric trap. Near the HAS, the suppression of radiative channels reduces the usual contributions and brings out nonlocal components associated with higher-order field gradients and generalized polarizabilities. These terms reshape the topology of the force landscapes, producing off-axis stable equilibria, two-point trapping, and extended weak-force regions, all of which are not attainable in standard scattering conditions.

*Conclusion.* – In this work, we have uncovered a previously unexplored regime of light-matter interaction, which fundamentally extends the paradigm of optomechanics. By introducing higher-order polarizability tensors that are linked to spatial gradients of the incident electromagnetic field we demonstrated that optical forces acting on resonant high-index nanoparticles can be governed not by the field itself, but by its higher-order spatial derivatives. The resulting previously neglected nonlocal optomechanical response can become dominant, underpinning peculiar particle dynamics even under trivial illumination.

We identified the hybrid anapole state as one of the possible essential prerequisites for observing nonlocal optomechanical effects. Due to near-zero far-field scattering and strong internal field localization the hybrid anapole state eliminates competing linear multipolar channels, allowing nonlocal gradient-driven components of multipoles to shape the optomechanical response. We showed that for a nanoparticle supporting the hybrid anapole state its interaction with the incident beam strongly depends on the particle size, giving rise to five distinct optomechanical effects: classical trapping, saddle-shaped bending of trajectories, asymmetric repulsion, asymmetric trapping with zero-force plateau and trapping with two equilibrium positions. Moreover, the last two regimes emerge exclusively due to the nonlocal terms and have no analogues in conventional optomechanics.

Therefore, special scattering regimes, such as, e.g., hybrid anapole, serve as a gateway to a new class of optomechanical phenomena, where the true nonlocal nature of electromagnetic forces becomes accessible. Our findings lay the foundation for advanced optical tweezing concepts, next-generation lab-on-a-chip devices, and novel architectures for resonant optomechanical control across physics, chemistry, and biology.

*Acknowledgements*

This work was supported by the Russian Science Foundation № 25-22-00266.

*Author Contributions*

These authors contributed equally: Susanna R. Rozental, Denis A. Kislov.

# Supplementary Material for

# **Nonlocal Optomechanics: Hybrid Anapole Opens a New Route to Optical Tweezing**

Susanna R. Rozental[1], Denis A. Kislov[1,3], Ilia M. Fradkin[1,3], Nikita S. Babich[2], Vasiliy Fedotov[3], Sergey Novikov[1], Vjaceslavs Bobrovs[4], Shangran Xie[5], Oleg Minin[6], Igor Minin[6], Lei Gao[7], Yu-Ling Wu[8], Lei Gong[8], Alexey Bolshakov[1], Alexey Arsenin[9], Alexander S. Shalin[1,3,7,10]

[1] Moscow Center for Advanced Studies, Kulakova str. 20, Moscow 123592, Russia
[2] ITMO University, Kronverkskii 49, St. Petersburg 197198, Russia
[3] Skolkovo Institute of Science and Technology, Bolshoi Boulevard 30, 121205 Moscow, Russia
[4] Riga Technical University, Institute of Photonics, Electronics and Telecommunications, 1048 Azenes street 12, Riga, LV 1658
[5] Beijing Institute of Technology, Beijing 100081, China
[6] Tomsk Polytechnic University, Lenina 36, Tomsk, 634050, Russia
[7] School of Optical and Electronic Information, Suzhou City University, Suzhou 215104, China
[8] Department of Optics and Optical Engineering, University of Science and Technology of China, Hefei 230026, China
[9] Emerging Technologies Research Center, XPANCEO, Internet City, Emmay Tower, Dubai, United Arab Emirates
[10] Faculty of Physics, M. V. Lomonosov Moscow State University, Moscow 119991, Russia

# Contents

# 1 Methods

## 1.1 Simulation

To simulation we used COMSOL Multiphysics®, module 'Electromagnetic Waves, Frequency Domain'. A cylindrical silicon nanoparticle with varying radius R and height H was placed into the cylindrical shaped computational domain with dimensions proportional to the nanoparticle size. Perfectly Matched Layers (PML) with the thickness of $\lambda/2$ were used to simulate the escape of a scattered light to infinity. A virtual cylindrical surface with radius of $R_1 = 1.5R$ and a height of $H_1 = 1.5H$, surrounding the nanoparticle, was additionally introduced for integrating the Maxwell stress tensor in the calculation of optical forces. To simulate laser radiation, we used a Gaussian beam in the approximation of plane waves, the beam width in the focus was equal to the wavelength ($w_0 = \lambda$). For the calculations, the entire domain was filled with a tetrahedral grid of variable size. The inner region of the particle was determined with a maximum grid size of $\lambda/30$, while the integration surface had a maximum of $\lambda/35$. In the environment, the maximum size of mesh was equal to $\lambda/5$. This grid is sufficient to accurately simulate fields in the whole domain.

## 1.2 Multipole decomposition

In this work we used two different multipole decompositions: Cartesian – to demonstrate the fulfillment of a hybrid anapole state condition and spherical to calculate scattering cross section and optical forces. All expressions for the multipoles are given in Section 1 of Supplementary.

## 1.3 Gaussian beam simulation

To calculate the field of a Gaussian beam linearly-polarized along the x-axis and propagating along the z-axis, we use the paraxial approximation, which is given by [1]:

$$\begin{cases} E^x_{gauss} \approx E_0 \dfrac{e^{-\rho/(1+z^2/z_0^2)}}{\sqrt{1+z^2/z_0^2}} e^{i\{kz[1+z_0\rho^2/k(z^2+z_0^2)]-arctg(z/z_0)\}}, \\ E^y_{gauss} = 0, \\ E^z_{gauss} \approx -\dfrac{ix}{z_0} E_x \dfrac{e^{-i \cdot arctg(z/z_0)}}{\sqrt{1+z^2/z_0^2}}, \\ k = \dfrac{\omega}{c}; \; \rho = \dfrac{\sqrt{x^2+y^2}}{w_0}; \; z_0 = \dfrac{kw_0^2}{2}, \end{cases} \quad (5)$$

Where $c$ is the speed of light in vacuum, $k$ is the wavenumber of light, $w_0$ is the characteristic radius of the beam at its waist (focus) and $z_0$ is the Rayleigh range.

It is important to note that a Gaussian beam has a purely real x-component and a purely imaginary z-component. This is a fundamental property of the Gaussian beam (which is described in more detail, for example, in [2]).

*1.4 Langevin dynamics*

To study the planar dynamics of nanoparticles in optical tweezers, we utilize the Langevin equation, which, in general, has the form:

$$m\frac{d^2\mathbf{r}(t)}{dt} = \mathbf{F}_{tr}(\mathbf{r},t) - \gamma\frac{d\mathbf{r}(t)}{dt} + \mathbf{f}_{rand}(t), \tag{6}$$

where $m$ is the mass of the nanocylinder, $\mathbf{r}$ is the location of the mass center, $\mathbf{F}_{tr} = \mathbf{F}_x + \mathbf{F}_y$ is the total transverse optical force, $\gamma$ is a damping factor due to viscous friction of the surrounding medium, and $\mathbf{f}_{rand}$ is a stochastic force. The medium is assumed to be air, which allows us to neglect the impact of viscous and stochastic forces and consider only the effects of optical forces.

The results of applying this approach to nanoparticles of various radii are shown in the main text, Fig. 4, right panels a)-e).

*1.5 Optical potential calculation*

An optomechanical interaction can also be characterized using a potential energy approach; for spherical nanoparticles, this is described in detail in [3]. For a cylindrical nanoparticle, however, the optical force may include non-conservative components. Nevertheless, we can define an energy-dimension quantity $U$ representing the work needed to move the particle from infinity to a given point against the optical forces. In this study, we are interested in the transverse optical potential, defined as the work against the transverse components $F_x$ and $F_y$. As the force distribution is computed numerically on a discrete grid, the integral is approximated as follows:

$$U(x_0, y_0) = -\int_{\infty}^{(x_0,y_0)} \mathbf{F}_{opt} \cdot d\mathbf{r} \approx -\iint_S F_{opt}(x,y)dxdy \tag{7}$$

Where $S = [0, L/2] \times [0, L/2]$ is a square integration domain with side $L = 2\mu m$ length. The grid was uniform with a step $\Delta x = \Delta y = 100 nm$, N = 20 is the number of grid nodes along one axis.

In the numerical calculation, we used the previously calculated matrices of optical force values along x and y axis $\{F^x_{i,j}\}$ and $\{F^y_{i,j}\}$. The energy at the origin was taken to be zero. Then, the energy values were calculated along the x and y axes. The values in the entire domain were then calculated iteratively by averaging the values calculated along the two paths (along x and y). The final computational scheme for energy:

$$\begin{cases} U_{1,1} = U(0,0) = 0 \\ U_{i,1} = U_{i-1,1} - F^x_{i,j}\Delta x, \quad i = 2...N \\ U_{1,j} = U_{1,j-1} - F^y_{i,j}\Delta y, \quad j = 2...N \\ U_{i,j} = \frac{1}{2}\left[(U_{i-1,j} - F^x_{i,j}\Delta x) + (U_{i,j-1} - F^y_{i,j}\Delta y)\right], \quad i,j = 2...N \end{cases} \quad (8)$$

The grid was uniform with a step $\Delta x = \Delta y = 100 nm$, N = 20 is the number of grid nodes along one axis, $\{F^x_{i,j}\}$ and $\{F^y_{i,j}\}$ are matrices of optical force values along x and y axes.

## 2  Hybrid anapole state

### 2.1  Multipole decompositions

To describe the electromagnetic fields scattered by nanoparticles, we used two types of multipole decomposition: exact spherical multipoles (were used to most of the calculations of scattered fields and optical forces) and Cartesian multipoles with toroidal terms (were used to proof the existence of a hybrid anapole state).

Let us write down firs four orders of spherical multipoles up to octupoles, expressed in Cartesian coordinates. The necessity of using octupoles for the study of HAS and will be discussed later in the section 3.4. At the same time, taking into account the multipoles of higher orders are not unnecessary, since without them there is a good convergence of the forces with those calculated through the Maxwell tensor describing the full field. Exact expressions for dipoles and quadrupoles are given in [4], for octupoles in [5,6]:

**Table 1. First four orders of the spherical multipoles in Cartesian coordinates**

| Multipole type | Expression |
|---|---|
| Electric dipole | $p_\alpha = \dfrac{i}{\omega}\left\{\int J_\alpha^\omega j_0(kr)d^3\mathbf{r} + \dfrac{k^2}{2}\int\left[3(\mathbf{r}\cdot\mathbf{J}^\omega)r_\alpha - r^2 J_\alpha^\omega\right]\dfrac{j_2(kr)}{(kr)^2}d^3\mathbf{r}\right\}$ |
| Magnetic dipole | $m_\alpha = \dfrac{3}{2}\int[\mathbf{r}\times\mathbf{J}^\omega]_\alpha \dfrac{j_1(kr)}{kr}d^3\mathbf{r}$ |
| Electric quadrupole | $EQ_{\alpha\beta} = \dfrac{3i}{\omega}\left\{\int\left[3(r_\beta J_\alpha^\omega + r_\alpha J_\beta^\omega) - 2(\mathbf{r}\cdot\mathbf{J}^\omega)\delta_{\alpha\beta}\right]\dfrac{j_1(kr)}{kr}d^3\mathbf{r} +\right.$ $\left. +2k^2\int\left[5r_\alpha r_\beta(\mathbf{r}\cdot\mathbf{J}^\omega) - (r_\beta J_\alpha^\omega + r_\alpha J_\beta^\omega)r^2 - r^2(\mathbf{r}\cdot\mathbf{J}^\omega)\delta_{\alpha\beta}\right]\dfrac{j_3(kr)}{(kr)^3}d^3\mathbf{r}\right\}$ |
| Magnetic quadrupole | $MQ_{\alpha\beta} = 15\int\left\{r_\alpha[\mathbf{r}\times\mathbf{J}^\omega]_\beta + r_\beta[\mathbf{r}\times\mathbf{J}^\omega]_\alpha\right\}\dfrac{j_2(kr)}{(kr)^2}d^3\mathbf{r}$ |
| Electric octupole | $EO_{\alpha\beta\gamma} = \dfrac{15i}{\omega}\left\{\int r_\alpha r_\beta J_\gamma + r_\alpha r_\gamma J_\beta + r_\beta r_\gamma J_\alpha - \dfrac{1}{5}\left(\delta_{\beta\gamma}(2r_\alpha(\mathbf{r}\cdot\mathbf{J}) + r^2 J_\alpha) +\right.\right.$ $\left.\left. +\delta_{\alpha\gamma}(2r_\beta(\mathbf{r}\cdot\mathbf{J}) + r^2 J_\beta) + \delta_{\alpha\beta}(2r_\gamma(\mathbf{r}\cdot\mathbf{J}) + r^2 J_\gamma)\right)\dfrac{j_2(kr)}{(kr)^2}d^3\mathbf{r}\right\}$ |
| Magnetic octupole | $MO_{\alpha\beta\gamma} = \dfrac{105}{4}\left\{\int r_\alpha r_\beta[\mathbf{r}\times\mathbf{J}]_\gamma + r_\alpha r_\gamma[\mathbf{r}\times\mathbf{J}]_{\gamma\beta} + r_\beta r_\gamma[\mathbf{r}\times\mathbf{J}]_\alpha -\right.$ $\left. -\dfrac{r^2}{5}\left[\delta_{\alpha\beta}[\mathbf{r}\times\mathbf{J}]_\gamma + \delta_{\alpha\gamma}[\mathbf{r}\times\mathbf{J}]_\beta + \delta_{\beta\gamma}[\mathbf{r}\times\mathbf{J}]_\alpha\right]\dfrac{j_3(kr)}{(kr)^3}d^3\mathbf{r}\right\}$ |

Where all integrals are taken over the volume of the particle, indices $\alpha, \beta, \gamma$ takes values from a set $\{x, y, z\}$, $\mathbf{J}^\omega = \mathbf{J}^\omega(\mathbf{r})$ is the spatially localized electric current density distribution corresponding to light with angular frequency $\omega$, $k = \frac{2\pi}{\lambda} = \frac{\omega}{c}$ is the wavenumber of light, $j_n(kr)$ is a spherical Bessel function of the n-th order.

We can also use the decomposition of fields into Cartesian moments (also called irreducible) and toroidal terms, which are higher order summands in the Taylor series expansion of scalar and vector field potentials [7]. Under the toroidal terms we will understand both toroidal multipoles and mean square radii of various orders, a more detailed classification of which is given in [8]:

**Table 2. First three orders of irreducible multipoles**

| Multipole type | Expression |
|---|---|
| Electric dipole | $p_\alpha^c = \frac{i}{\omega} \int J_\alpha^\omega d^3\mathbf{r}$ |
| Magnetic dipole | $m_\alpha^c = \frac{1}{2} \int [\mathbf{r} \times \mathbf{J}^\omega]_\alpha d^3\mathbf{r}$ |
| Electric quadrupole | $EQ_{\alpha\beta}^c = \frac{i}{\omega} \left\{ \int \left[ 3(r_\beta J_\alpha^\omega + r_\alpha J_\beta^\omega) - 2(\mathbf{r} \cdot \mathbf{J}^\omega)\delta_{\alpha\beta} \right] d^3\mathbf{r} \right.$ |
| Magnetic quadrupole | $MQ_{\alpha\beta}^c = \int \left\{ r_\alpha [\mathbf{r} \times \mathbf{J}^\omega]_\beta + r_\beta [\mathbf{r} \times \mathbf{J}^\omega]_\alpha \right\} d^3\mathbf{r}$ |

**Table 3. Toroidal terms corresponding to irreducible multipoles**

| Multipole type | Expression |
|---|---|
| Electric dipole | $T_\alpha^p = \frac{1}{10} \int \left[ (\mathbf{J}^\omega \cdot \mathbf{r})r_\alpha - 2r^2 J_\alpha^\omega \right] d^3\mathbf{r}$ ; <br> $T_\alpha^{p\langle 2\rangle} = \frac{1}{280} \int \left[ 3r^4 J_\alpha^\omega - 2r^2(\mathbf{J}^\omega \cdot \mathbf{r}) J_\alpha^\omega \right] d^3\mathbf{r}$ |
| Magnetic dipole | $T_\alpha^m = \frac{i\omega}{20} \int r^2 [\mathbf{r} \times \mathbf{J}^\omega]_\alpha d^3\mathbf{r}$ |
| Electric quadrupole | $T_{\alpha\beta}^{EQ} = \frac{1}{14} \int \left[ 4(\mathbf{J}^\omega \cdot \mathbf{r})r_\alpha r_\beta + 2\delta_{\alpha\beta} r^2 (\mathbf{J}^\omega \cdot \mathbf{r}) - 5r^2(r_\beta J_\alpha^\omega + r_\alpha J_\beta^\omega) \right] d^3\mathbf{r}$ |
| Magnetic quadrupole | $T_{\alpha\beta}^{MQ} = \frac{i\omega}{42} \int r^2 \left\{ r_\alpha [\mathbf{r} \times \mathbf{J}^\omega]_\beta + r_\beta [\mathbf{r} \times \mathbf{J}^\omega]_\alpha \right\} d^3\mathbf{r}$ |

The sum of each irreducible multipole with its toroidal term is approximately equal to the corresponding spherical multipole. Therefore, both decompositions can be employed to describe the scattered field. The total scattering cross section of nanoparticle in terms of spherical multipole moments up to the octupoles [9] is given by:

$$\sigma_{sca} = \frac{k^2}{\pi\varepsilon_0^2 |\mathbf{E}_{inc}|^2}\left[\frac{1}{6}\sum_{\alpha=x,y,z}\left(|p_\alpha|^2 + \left|\frac{m_\alpha}{c}\right|^2\right) + \frac{1}{720}\sum_{\alpha,\beta=x,y,z}\left(|kQ^e_{\alpha\beta}|^2 + \left|\frac{kQ^m_{\alpha\beta}}{c}\right|^2\right) + \right.$$
$$\left. + \frac{1}{315}\sum_{\alpha,\beta,\gamma}\left(|k^2 O^e_{\alpha\beta\gamma}|^2 + \left|\frac{k^2 O^m_{\alpha\beta\gamma}}{c}\right|^2\right)\right] \quad (9)$$

Where $p_\alpha$, $m_\alpha$ are the electric and magnetic dipole moments, respectively; $Q^e_{\alpha\beta}$, $Q^m_{\alpha\beta}$ are the electric and magnetic quadrupole moments; $O^e_{\alpha\beta\gamma}$, $O^m_{\alpha\beta\gamma}$ are the electric and magnetic octupole moments. $|\mathbf{E}_{inc}|$ is the electric field amplitude of the incident wave, $k$ is the wavenumber in vacuum, $\varepsilon_0$ is the vacuum permittivity and $c$ is the speed of light.

Furthermore, the scattering cross-section can be described in terms of Cartesian multipoles and toroidal terms up to quadrupoles is given by [8]:

$$\sigma_{sca} \approx \frac{k^2}{\pi\varepsilon_0^2 |\mathbf{E}_{inc}|^2}\left[\frac{1}{6}\sum_{\alpha=x,y,z}\left(\left|p^c_\alpha + \frac{ik}{c}T^p_\alpha + \frac{ik^3}{c}T^{p\langle 2\rangle}_\alpha\right|^2 + \frac{1}{c}\left|m^c_\alpha + \frac{ik}{c}T^m_\alpha\right|^2\right) + \right.$$
$$\left. + \frac{1}{720}\sum_{\alpha,\beta=x,y,z}\left(\left|EQ^c_\alpha + \frac{ik}{c}T^{EQ}_\alpha\right|^2 + \frac{1}{c^2}\left|MQ^c_\alpha + \frac{ik}{c}T^{MQ}_\alpha\right|^2\right) + ...\right] \quad (10)$$

To calculate the total scattering cross section, it is also necessary to take into account the contribution of octupole moments, which can be calculated using spherical multipoles as shown in formula (9).

## 2.2 HAS condition

In light of this formalism, an anapole state can be defined as a destructive interference of the irreducible multipole with the corresponding toroidal multipole, which results in the cancellation of the corresponding contribution to the scattering cross-section. For example, the scattering cross section of electric dipole is proportional to the sum of cartesian multipole and toroidal terms:

$$\sigma_{sca}^{p} = \frac{k^2}{6\pi\varepsilon_0^2 |\mathbf{E}_{inc}|^2} \left| \mathbf{p}_{car} + \frac{ik}{c}\mathbf{T}^p + \frac{ik^3}{c}\mathbf{T}^{p\langle 2\rangle} \right|^2 = 0 \quad (11)$$

Therefore, the condition of electric dipole anapole can be formulated as [10]:

$$\mathbf{p}_{car} = -\left( \frac{ik}{c}\mathbf{T}^p + \frac{ik^3}{c}\mathbf{T}^{p\langle 2\rangle} \right) \quad (12)$$

An anapole condition requires the irreducible dipole and its toroidal counterpart to possess equal magnitudes and a phase difference of $\Delta\varphi = \pm\pi$. Recently, it was shown that broken spherical symmetry allows the spectral overlap of up to four such conditions, creating a "hybrid anapole" with near-zero scattering [11,12]. We confirm these destructive interference conditions for the first two multipole orders in Figure S1, which plots the magnitudes and phase differences calculated from the expressions in Table 2 and Table 3.

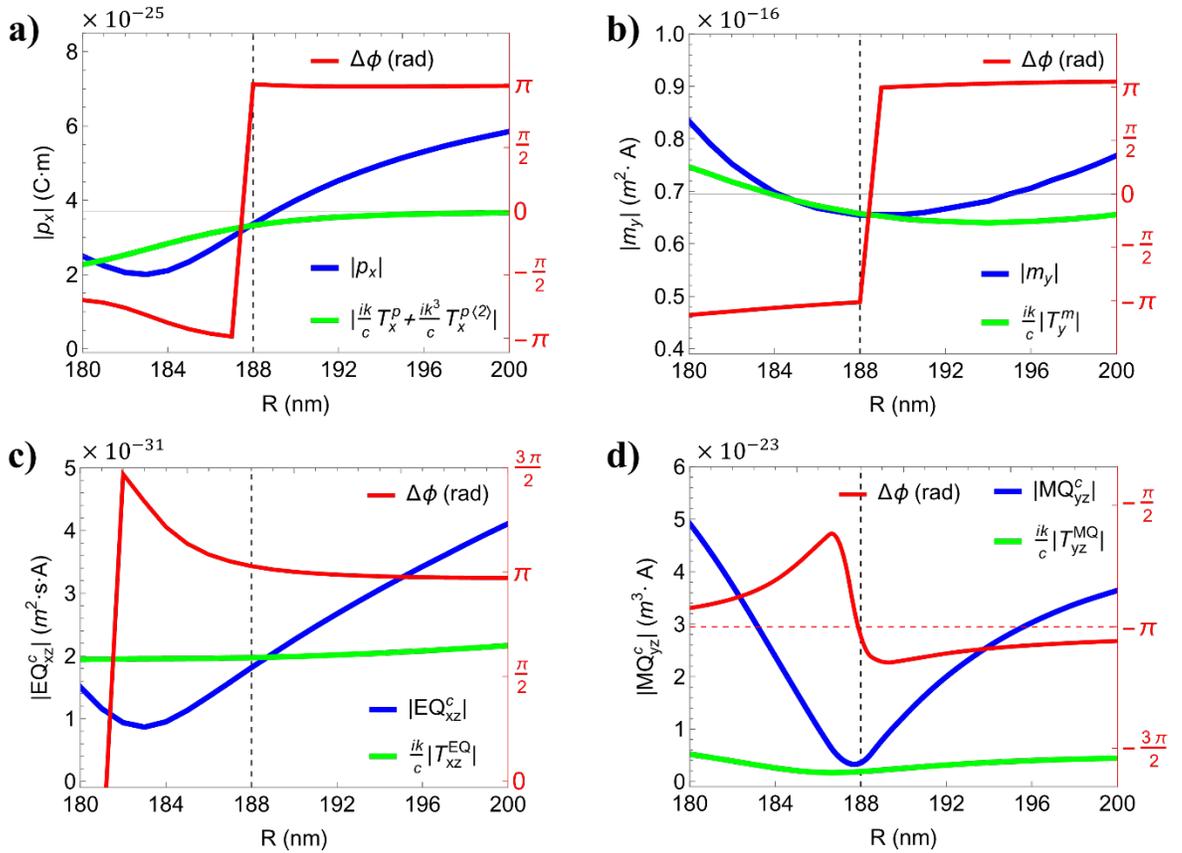

Figure S1. Amplitudes and phase differences between the cartesian multipoles and their toroidal counterparts. a) the basic electric and electric toroidal dipoles; b) the basic magnetic and magnetic toroidal dipoles; c) the basic electric and electric toroidal quadrupoles; d) the basic magnetic and magnetic toroidal quadrupoles. Amplitudes correspond to the left ordinate-axis, and phase

differences are read from the right ordinate-axis. The black dashed line is drawn through the minimum of the total scattering cross section (corresponding to R = 188 nm).

Figure S1 shows, that for a nanoparticle with R = 188 nm the Cartesian multipoles and their toroidal part have the same magnitudes, and the phase difference between them ±π for both electric and magnetic dipoles and quadrupoles. This means that the anapole condition is satisfied for these multipoles simultaneously, resulting in a hybrid anapole state.

## 3 Optical forces

### 3.1 General formulation

The average optical force acting on the particle can be calculated by integrating Maxwell stress tensor on a closed surface surrounding the object:

$$\langle \mathbf{F} \rangle_t = \oiint_{\partial V} \langle \vec{\mathbf{T}}(\mathbf{r},t) \rangle \cdot \mathbf{n}(\mathbf{r}) da = \frac{1}{2} \text{Re} \oiint_{\partial V} \vec{\mathbf{T}}(\mathbf{r},t) \cdot \mathbf{n}(\mathbf{r}) da \qquad (13)$$

Where $\langle \cdot \rangle_t$ means the time average over an oscillation period of the electromagnetic field, $\mathbf{n}$ is the normal unit vector to the surrounding surface $\partial V$, is $da$ is an element of the surface, $\vec{\mathbf{T}}$ is the average Maxwell's stress tensor (MST) in vacuum, which is given by:

$$\vec{\mathbf{T}} = \varepsilon \varepsilon_0 \mathbf{E} \otimes \mathbf{E}^* + \mu \mu_0 \mathbf{H} \otimes \mathbf{H}^* - \frac{1}{2} \left( \varepsilon \varepsilon_0 \mathbf{E} \cdot \mathbf{E}^* + \mu \mu_0 \mathbf{H} \cdot \mathbf{H}^* \right) \vec{\mathbf{I}} \qquad (14)$$

Where $\otimes$ denotes a dyadic operation over two tensors, $\vec{\mathbf{I}}$ denotes the unit tensor, $\mathbf{E}$ and $\mathbf{H}$ are the total fields, defined as the sum of the incident and scattered fields, given by:

$$\mathbf{E} = \mathbf{E}_{inc} + \mathbf{E}_{sca}; \quad \mathbf{H} = \mathbf{H}_{inc} + \mathbf{H}_{sca} \qquad (15)$$

The time-averaged optical force on a dielectric particle could be expressed in terms of the induced multipoles:

$$F = [F_p + F_{Q^e} + F_{O^e} + \cdots] + [F_m + F_{Q^m} + F_{O^m} + \cdots] + \\ + [F_{pm} + F_{Q^e Q^m} + F_{O^e O^m} + \cdots] + [F_{pQ^e} + F_{Q^e O^e} + \cdots] + [F_{mQ^m} + F_{Q^m O^m} + \cdots] \qquad (16)$$

Then, the i-th component of the optical force is given by [13]:

$$F_i = \frac{1}{2}\text{Re}[p_j\nabla_i E_j^*] + \frac{1}{12}\text{Re}[Q_{jk}^e\nabla_i\nabla_k E_j^*] + \frac{1}{12}\text{Re}[O_{jkl}^e\nabla_i\nabla_l\nabla_k E_j^*] +$$

$$+ \frac{1}{2}\text{Re}[m_j\nabla_i B_j^*] + \frac{1}{12}\text{Re}[Q_{jk}^m\nabla_i\nabla_k B_j^*] + \frac{1}{12}\text{Re}[O_{jkl}^m\nabla_i\nabla_l\nabla_k B_j^*] -$$

$$- \frac{k^4}{12\pi\varepsilon_0 c}\text{Re}[\varepsilon_{ijk}p_j m_k^*] - \frac{k^6}{2160\pi\varepsilon_0 c^2}\text{Re}[\varepsilon_{ijk}Q_{lj}^e(Q_{lk}^m)^*] - \frac{k^8}{7560\pi\varepsilon_0 c}\text{Re}[\varepsilon_{ijk}O_{lnj}^e(O_{lnk}^m)^*] - \quad (17)$$

$$- \frac{k^5}{120\pi\varepsilon_0}\text{Im}[Q_{ij}^e p_j^*] - \frac{k^7}{1890\pi\varepsilon_0}\text{Im}[O_{ijk}^e(Q_{jk}^e)^*] -$$

$$- \frac{k^5}{120\pi\varepsilon_0 c^2}\text{Im}[Q_{ij}^m m_j^*] - \frac{k^7}{1890\pi\varepsilon_0 c}\text{Im}[O_{ijk}^m(Q_{jk}^m)^*]$$

Where $\varepsilon_{ijk}$ is Levi-Civita symbol, i, j, k, l $\in$ {x,y,z}. According to Einstein notation, the summation is performed over all duplicated indices except for the free index i.

## 3.2 Transverse forces

Let us compare transverse forces at different nanoparticle radii: $F_x$ arising at displacement of the particle along the x-axis, and $F_y$ arising at displacement of the particle along the y-axis – Figure S2.

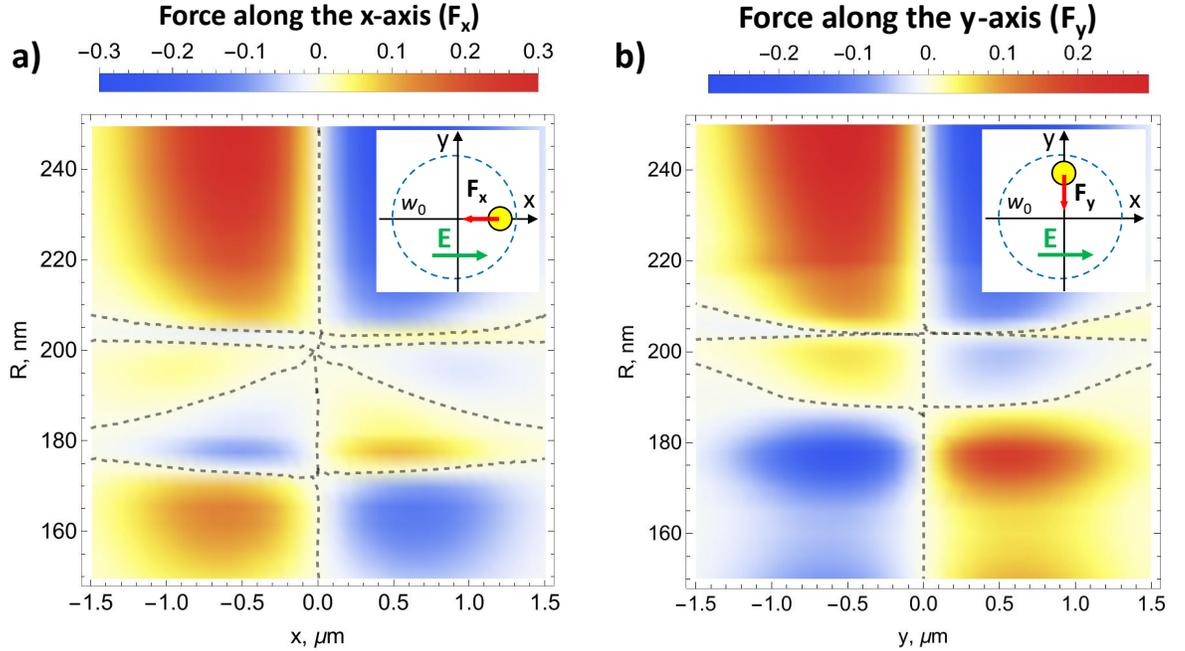

Figure S2. Optical forces in the radius-coordinate space: a) $F_x$ in (x, R) space, b) $F_y$ in (y, R) space (dotted lines indicate the areas of zero forces, inserts show the direction of optical force components relative to the polarization of incident light).

Figure S2 shows that near the HAS the behavior of the force along the x-axis differs from the behavior of the force along the y-axis. This is likely due to the linear polarization of the incident radiation along the x-axis, which leads to a radial asymmetry of scattering and optomechanical effects. The combination of different forces along different axes in the focal plane results in a unique distribution of the total transverse force. To investigate these effects, we selected 5 different radii of nanoparticles, corresponding to different optomechanical regimes: 188 nm (HAS), 170 nm, 180 nm, 198 nm and 210 nm. The forces at these radii are shown in Fig. 3 in the main text.

### 3.3 Pressure force and "HAS destruction" effect

In this section, we examine the effect of the HAS on the light pressure ($F_z$ component of the total force). The results are presented in Figure S3, where the dependence of the pressure force in the center of the beam on the radius of the nanocylinder is shown together with the spatial distributions of optical pressure for specific nanoparticle radii.

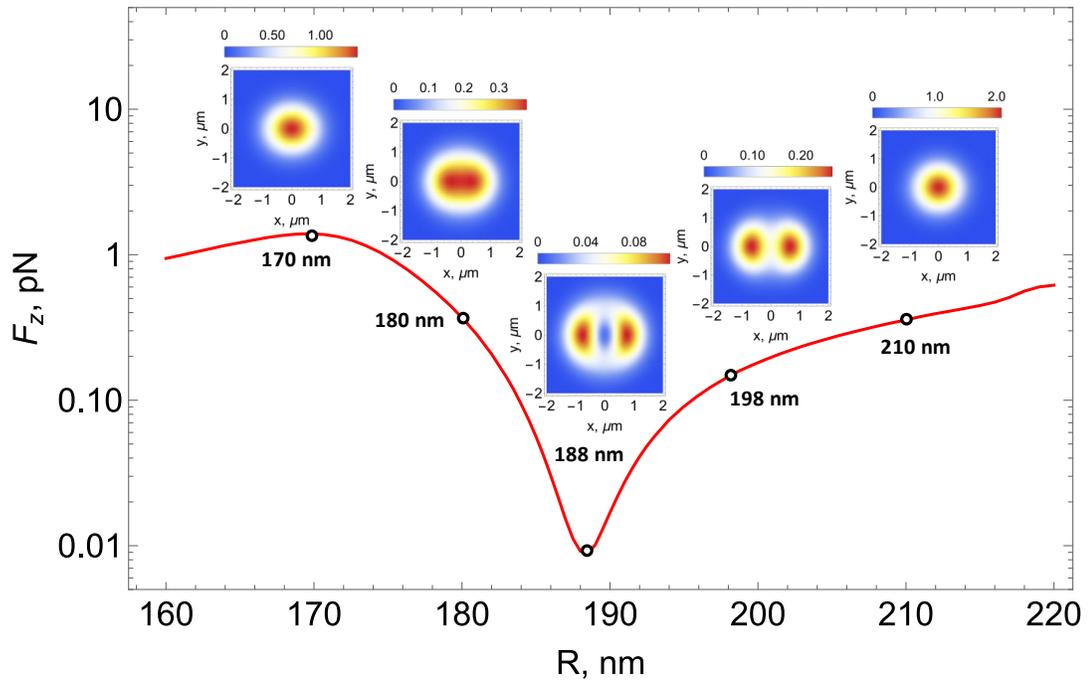

Figure S3. Light pressure, Fz, acting on a nanocylinder located in the centre of the focal plane versus nanocylinder radius. Insets show the spatial distribution of Fz in the focal plane for selected R marked on the red curve.

Noticeable, that in regimes that are far from the hybrid anapole (e.g. R = 170 nm and 210 nm), light pressure assumes a radially symmetric Gaussian distribution in the focal plane. As the hybrid anapole regime is approached, the symmetry is gradually broken, the force maximum expands along the x-axis and eventually the distribution decays into 2 lobes in the HAS (R = 188 nm). More specifically, the nanocylinder experiences the lowest pressure on the axis of the beam, but as it drifts away from the axis, the pressure starts to increase and the increase is particularly pronounced along the x-direction, which is the consequence of the beam being linearly polarized along the x-axis. At the edge of the beam, however, the pressure falls and, following the exponential decrease of light intensity, drops to the values characteristic of the normal scattering regime. We termed the described effect as "HAS destruction". This effect is conditioned by a response of multipole moments near the HAS on the high-order field gradients, which will be discussed in detail in Section 4.

*3.4 The effect of octupoles*

In the hybrid anapole state (HAS), octupole moments provide a dominant contribution to the scattered field. Figure S4 illustrates the difference between the scattering efficiency computed from dipole and quadrupole terms alone and the total scattering efficiency including the octupole terms.

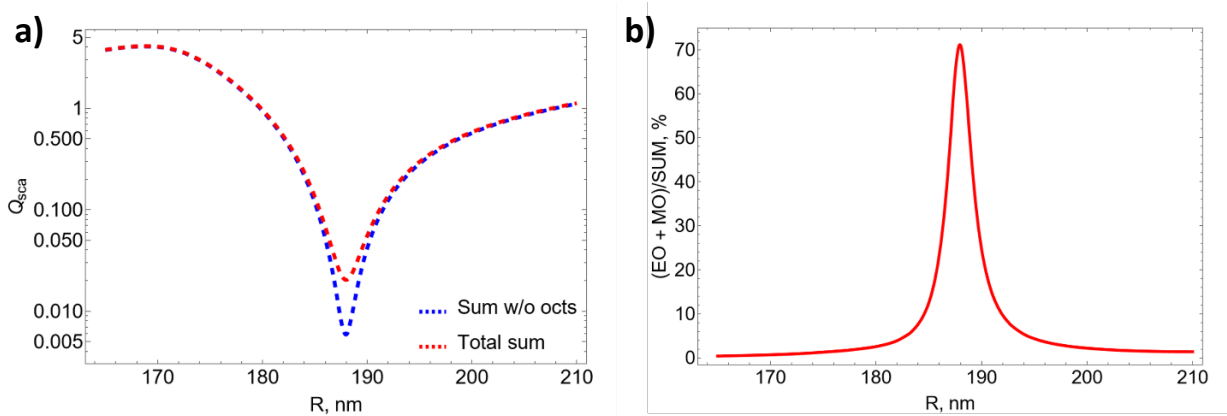

Figure S4. a) A comparison of the scattering efficiency calculated with octupole terms (red dashed curve) and without octupoles (blue dashed curve); b) Octupole (magnetic + electric) contribution to total scattering efficiency versus particle radius. The peak in the region of the HAS (R = 188 nm) corresponds to an increase of octupole contribution up to 70 %

For the particles in the scattering regime octupole moments are usually negligible in comparison to dipoles and quadrupoles. However, as seen in Fig. 2(c) in the main text, in the HAS all the dipole and quadrupole components decrease by two times in magnitude, while octupole components remains the same, rendering their contribution to the scattering cross section the most significant.

Octupoles also make a large contribution to the transverse forces - Figure S5.

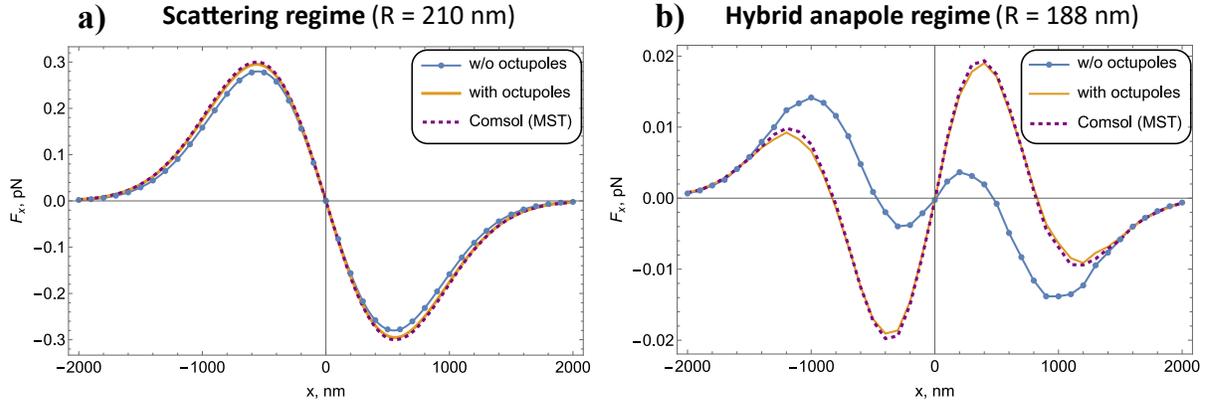

Figure S5. Comparison of total transverse force $F_x$, calculated using Maxwell stress tensor (pink dashed line) and using multipole decomposition up to octupoles (yellow line) with the force calculated without octupole terms (blue dotted line): a) in scattering regime (R = 210 nm), b) in the hybrid anapole state (R = 188 nm)

Figure S5(b) shows a significant difference between the force calculated using the multipoles up to octupoles (yellow line), and the truncated expansion without octupoles (blue dotted line). In the scattering regime Figure S5(a) this difference is negligible.

*3.5 Multipole components of force*

Using Eq. (17) we can calculate all the interceptional and recoil components of $F_x$ and $F_z$ forces and compare them in the HAS and in the conventional regime – Figure S6 and Figure S7.

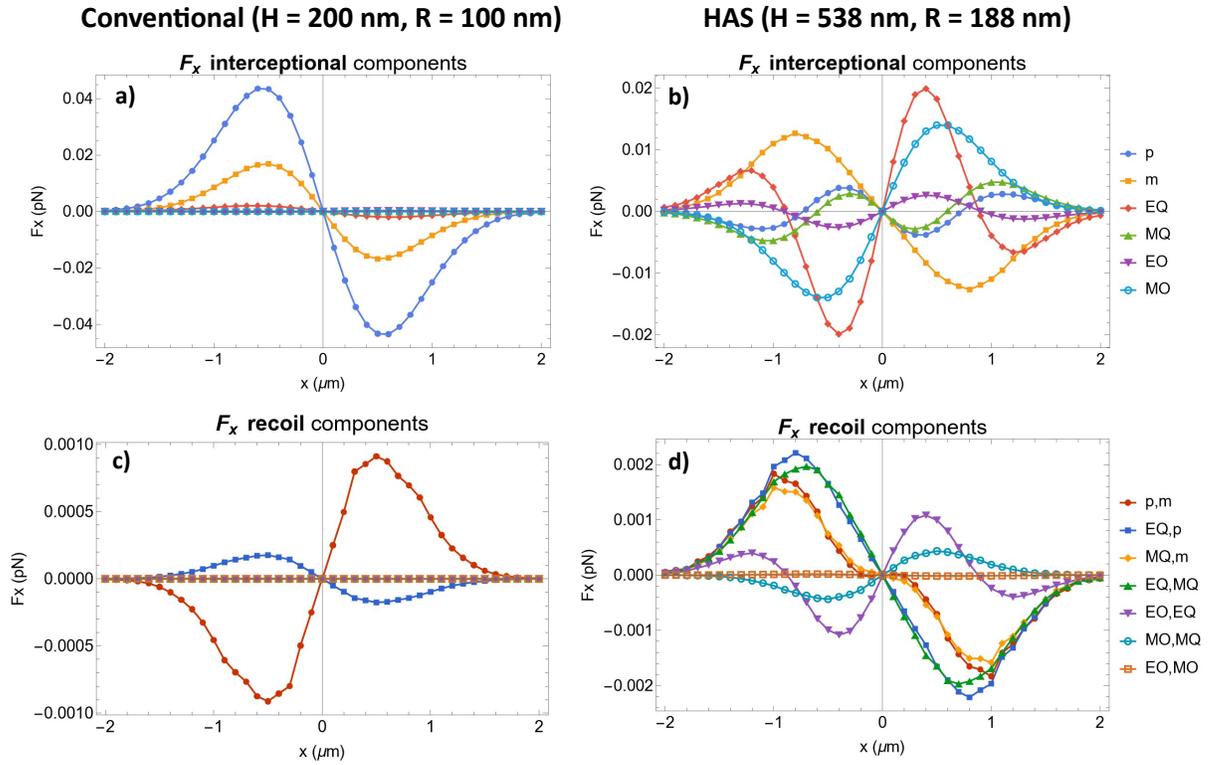

Figure S6. Interceptional and recoil components of the transverse force $F_x$ for the nanoparticle in the conventional scattering regime (a, c) and in the HAS (b, d).

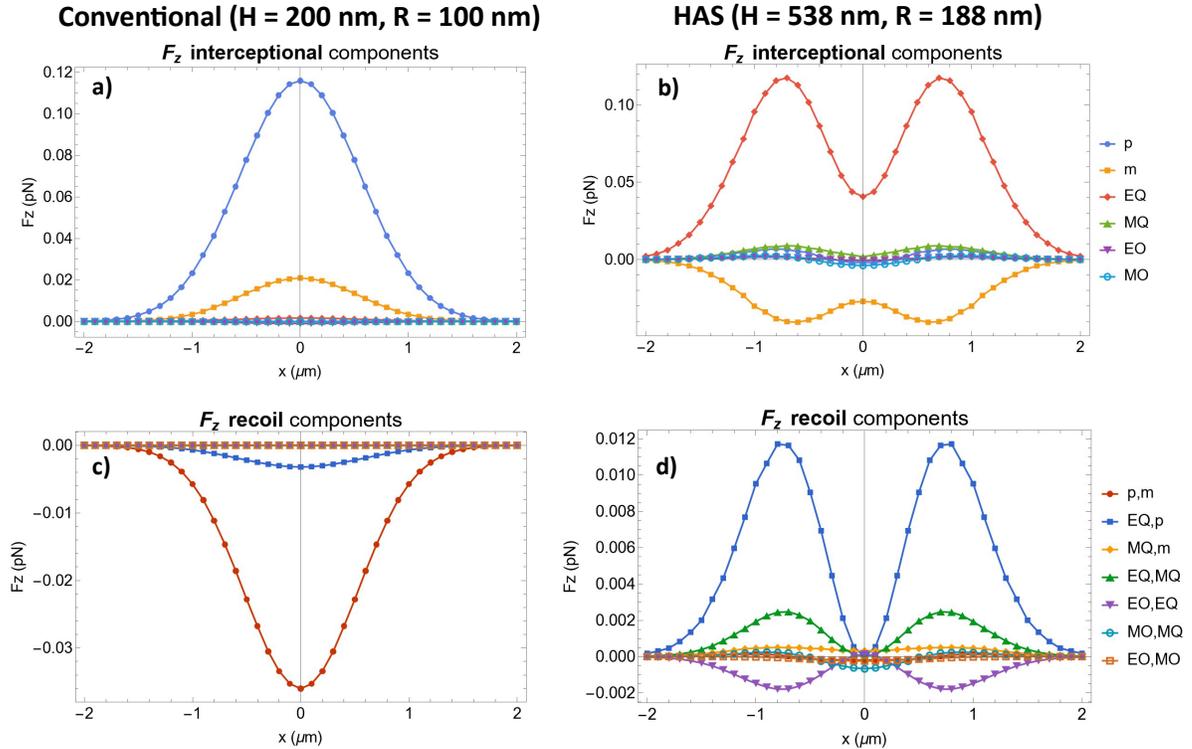

Figure S7. Interceptional and recoil components of the pressure force $F_z$ for the nanoparticle in the conventional scattering regime (a, c) and in the HAS (b, d).

Figure S6 shows that almost all the interceptional components in the HAS exhibit non-classical behavior with 2 zeroing points, while in the conventional regime the classical gradient force is observed. Recoil components in both cases are relatively small, but for nanoparticles with other parameters they can contribute significantly.

From Figure S7 we can see that the multipole components of the optical pressure force in the HAS have a dip at the beam axis, which is attributed to the so-called "HAS destruction effect", observed in Figure S3. This effect manifests itself in both dipole and quadrupole forces, rather than just in octupoles, as one might expect from the anapole conditions. The reason for this is the predominance of the response to higher derivatives of the fields in the HAS, which is discussed in the following section.

## 4  Nonlocal polarizabilities calculation

### 4.1  Theoretical background

To describe the difference in forces between hybrid anapole and non-anapole states, a simple and efficient method is required for determining the multipole moments of a particle in an arbitrary field, without performing rigorous calculations for each specific case. The most common approach involves introducing multipolar polarizability tensors, which relate the multipole moments of the particle to characteristics of the background field, $\mathbf{E}_{bg}$, acting upon it. These field characteristics are typically given by terms in an infinite series expansion. In most cases the first two terms of this series are enough for correct approximation of multipole moments of nanoparticle (it's a so-called "conventional" approximation). However, in case of a hybrid anapole state this approximation is not sufficient, as we show later in Section 4.4, so we need to consider higher-order terms to explain the unique force patterns.

This expansion can be constructed using a basis of functions that solve Maxwell's equations, such as spherical harmonics, or an arbitrary non-physical basis. Although the latter approach has its drawbacks, expanding the background field, for instance, in a Taylor series is extremely convenient for practical applications. This is because it is straightforward to calculate the electric field and its gradients at any point, unlike expansions in spherical harmonics. Without loss of generality, ley us consider the electric dipole moment:

$$\mathbf{p}(\mathbf{r}) = \mathbf{p}[\mathbf{E}_{bg}(\mathbf{r})] \qquad (18)$$

The background field can be expanded in a Taylor series as [14]:

$$E^{\beta}_{bg}(\mathbf{r}) = E^{\beta}_{bg}(0) + \frac{1}{1!} r_m \frac{\partial E^{\beta}_{bg}(0)}{\partial r_m} + \frac{1}{2!} r_m r_n \frac{\partial^2 E^{\beta}_{bg}(0)}{\partial r_n \partial r_m} + \ldots \qquad (19)$$

Due to the linearity of the functional with respect to the applied field, we write:

$$p^\alpha = p^\alpha[E_{bg}^\beta \mathbf{e}^\beta] + p^\alpha[r_m \frac{\partial E_{bg}^\beta}{\partial r_m}\mathbf{e}^\beta] + p^\alpha[\frac{1}{2}r_n r_m \frac{\partial^2 E_{bg}^\beta}{\partial r_n \partial r_m}\mathbf{e}^\beta] + ..., \quad (20)$$

where $\mathbf{e}^\beta$ is a unit vector along the $\beta$-axis. Applying the linearity property once more yields:

$$p^\alpha = p^\alpha[\mathbf{e}^\beta]E_{bg}^\beta + p^\alpha[r_m \mathbf{e}^\beta]\frac{\partial E_{bg}^\beta}{\partial r_m} + p^\alpha[\frac{1}{2}r_n r_m \mathbf{e}^\beta]\frac{\partial^2 E_{bg}^\beta}{\partial r_n \partial r_m} + ..., \quad (21)$$

This gives an expansion of the electric dipole moment in terms of the gradients of the background electric field. The coefficients of proportionality are the polarizability tensors, defined as:

$$\alpha_{\alpha\beta} = p^\alpha[\mathbf{e}^\beta], \quad \alpha_{\alpha m\beta} = p^\alpha[r_m \mathbf{e}^\beta], \quad \alpha_{\alpha mn\beta} = p^\alpha[\frac{1}{2}r_n r_m \mathbf{e}^\beta] \quad (22)$$

In other words, the corresponding polarizabilities can be calculated as the response to a uniform field, a linearly growing field with a unit gradient, and so on. However, none of these fields satisfy Maxwell's equations and thus cannot be directly applied to the particle in a numerical solver.

To address this, we employ the widely used scattered field formulation [15,16]. The main principle of this approach is that the field scattered by the particle field (defined as the difference between the total field and the background one) $\mathbf{E}_{sc}(\mathbf{r}) = \mathbf{E}_{tot}(\mathbf{r}) - \mathbf{E}_{bg}(\mathbf{r})$ can be determined as the radiation from an equivalent current source within the particle:

$$\mathbf{j}_{primary}(\mathbf{r}) = -\frac{i\omega}{4\pi}\Delta\varepsilon(\mathbf{r},\omega)\mathbf{E}_{bg}(\mathbf{r},\omega), \quad (23)$$

Where $\Delta\varepsilon = \varepsilon(\mathbf{r},\omega) - \varepsilon_{bg}(\mathbf{r},\omega)$ is the difference between the permittivity of the scatterer and the background media.

The total field is then the sum of the background and scattered fields:

$$\mathbf{E}_{tot}(\mathbf{r}) = \mathbf{E}_{bg}(\mathbf{r}) + \mathbf{E}_{sc}(\mathbf{r}) \quad (24)$$

and this total field is used to calculate the multipole moments.

Now, we can substitute any non-physical field into Eq. (23), and the corresponding currents might be used for further calculations of multipole moments without issue. While this might initially appear to be a trick for solving an otherwise intractable problem, it is not. The optical response to a non-physical field should only be interpreted within the context of an infinite series, which, when summed, represents the response to a physically realizable field. In practical applications, we always work with a finite, truncated series, which provides an approximate solution - this is both sufficient and intended for most use cases.

The general formula for the arbitrary multipole M is:

$$M^i[\mathbf{E}(\mathbf{r})] = M^i\left[\sum_{n=0}^{\infty}\sum_{k_1,k_2,k_3}\frac{\delta_{n,k_1+k_2+k_3}}{k_1!k_2!k_3!}\frac{\partial^k\left(\sum_{\beta=x,y,z}E_\beta \mathbf{e}_\beta\right)}{(\partial^{k_1}x)(\partial^{k_2}y)(\partial^{k_3}z)}x^{k_1}y^{k_2}z^{k_3}\right] =$$

$$= \sum_{n=0}^{\infty}\sum_{k_1,k_2,k_3}\frac{\delta_{n,k_1+k_2+k_3}}{k_1!k_2!k_3!}\sum_{\beta=x,y,z}M^i\left[x^{k_1}y^{k_2}z^{k_3}\mathbf{e}_\beta\right]\frac{\partial^k E_\beta}{(\partial^{k_1}x)(\partial^{k_2}y)(\partial^{k_3}z)} = \qquad (25)$$

$$= \sum_{n=0}^{\infty}\sum_{k_1,k_2,k_3}\frac{\delta_{n,k_1+k_2+k_3}}{k_1!k_2!k_3!}\sum_{\beta=x,y,z}\alpha^M_{ik_1k_2k_3\beta}\frac{\partial^k E_\beta}{(\partial^{k_1}x)(\partial^{k_2}y)(\partial^{k_3}z)},$$

Where $\mathbf{e}_\beta$ is a unit vector along the $\beta$-axis, the index set $i = (i_1...i_n)$ denotes the coordinates of the multipole (one index for a dipole, two for a quadrupole, etc.), $\alpha^M_{ik_1k_2k_3\beta}$ is a component of polarizability tensor corresponding to the response of the $i^{th}$ component of M to the gradients of β-th component of an incident field.

Now, the polarizability tensor of arbitrary rank in Eq. (25) is defined as a generalized response on power-law field of corresponding degree:

$$\alpha_{ik_1k_2k_3\beta} = M^i\left[x^{k_1}y^{k_2}z^{k_3}\mathbf{e}_\beta\right], \qquad (26)$$

Numerical simulations of response to power-law were conducted in COMSOL Multiphysics®, the geometry and parameter set were identical to the ones described in the in the main text.

## 4.2 Convergence of Taylor series

To find out which order of derivatives in the Taylor series is sufficient to approximate the full field, we plotted the shape of the Taylor-approximated field along the x-axis and compared it with the full Gaussian beam given by Eq. (5)

The field along the x-axis is affected only by the 2 and 4 derivatives in the Taylor series, so we can break the series on them and write down two approximate expressions:

$$\mathbf{E}_2(\mathbf{r}) = \mathbf{E}(0) + r_m \frac{\partial \mathbf{E}(0)}{\partial r_m} + r_m r_n \frac{\partial^2 \mathbf{E}(0)}{\partial r_n \partial r_m} \tag{27}$$

$$\mathbf{E}_4(\mathbf{r}) = \mathbf{E}(0) + r_m \frac{\partial \mathbf{E}(0)}{\partial r_m} + r_m r_n \frac{\partial^2 \mathbf{E}(0)}{\partial r_n \partial r_m} + r_m r_n r_l \frac{\partial^3 \mathbf{E}(0)}{\partial r_l \partial r_n \partial r_m} + r_m r_n r_l r_s \frac{\partial^4 \mathbf{E}(0)}{\partial r_s \partial r_l \partial r_n \partial r_m} \tag{28}$$

Now let us compare fields approximated up to 2 and up to 4 of the order of Taylor series ($\mathbf{E}_2(x)$ and $\mathbf{E}_4(x)$ respectively) with the exact field $\mathbf{E}_{gauss}(x) = \{E^x_{gauss}(x), 0, E^z_{gauss}\}$. Noteworthy, that $E^x_{gauss} = \mathrm{Re}[E^x_{gauss}]; E^z_{gauss} = \mathrm{Im}[E^z_{gauss}]$ and these expressions hold for all the spatial derivatives of these components.

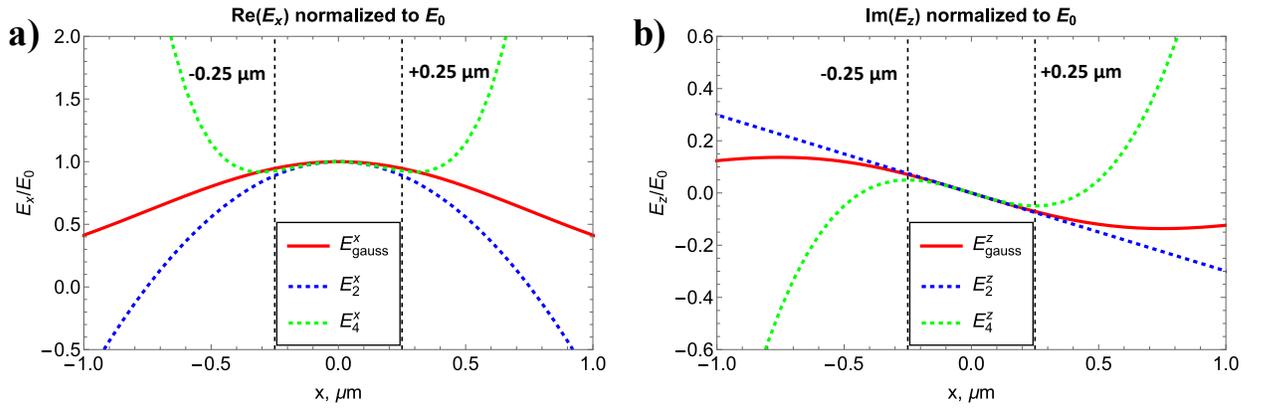

Figure S8. Dependence of the Gaussian beam profile, its 2nd and 4th derivatives normalized to the maximum on the x-coordinate: a) real part of the $E_x$, b) imaginary part of $E_z$.

It follows from Figure S8 that for a nanoparticle with a radius of 210 nm and smaller, the Taylor series with an accuracy of 4th order derivatives is sufficient to approximate the full field of a Gaussian beam inside the particle.

*4.3 Conventional approximation*

In the conventional approximation, which is widely used for low-index and small nanoparticles, multipole moments are given by the incident field at the particle's center multiplied by the corresponding polarizability tensors [14,17,18]:

$$\mathbf{p} = \alpha_p \mathbf{E}_{inc}; \quad \mathbf{m} = \alpha_m \mathbf{H}_{inc}$$
$$\mathbf{Q}^e = \alpha_{Q^e} \frac{\nabla \mathbf{E}_{inc} + \mathbf{E}_{inc} \nabla}{2}; \quad \mathbf{Q}^m = \alpha_{Q^m} \frac{\nabla \mathbf{H}_{inc} + \mathbf{H}_{inc} \nabla}{2}, \tag{29}$$

As in the Taylor series method, multipole moments and the electromagnetic field are related through polarizability tensors. However, in the conventional approximation magnetic moments are proportional to the magnetic field and its first gradient, while in Taylor series method all the multipoles are proportional only to the electric field gradients and its gradients. This difference could be overcome by using 3rd Maxwell's equation, which relates the magnetic field with the rotor to the electric field. Thus, the $i^{th}$ component of the magnetic field can be expressed in terms of the difference of the derivatives of the electric field of the form $(\partial_j E_k - \partial_k E_j)$, and after some symmetrization procedures, magnetic polarizability tensor of 2nd rank (representing response to a magnetic field) can be expressed in terms of electric polarizability tensors of 3rd rank (representing response to the 1st derivative of the electric field).

### 4.4 Electric dipole calculation

Now we can compare the multipole moments calculated using our method with the "conventional" approximation and the exact spherical multipoles from Table 1. For comparison, we chose the components of the electric dipole ($p_x$ and $p_z$), since in the conventional approximation electric dipole has the simplest formula for interpretation. More specifically, we examined $Re(p_x)$ and $Im(p_z)$, as these are the only components that determine the electric dipole force in our system. The comparison is made between the nanoparticle in the hybrid anapole state (H = 538 nm, R = 188 nm) and the nanoparticle with normal scattering (H = 200 nm, R = 100 nm) – Figure S9.

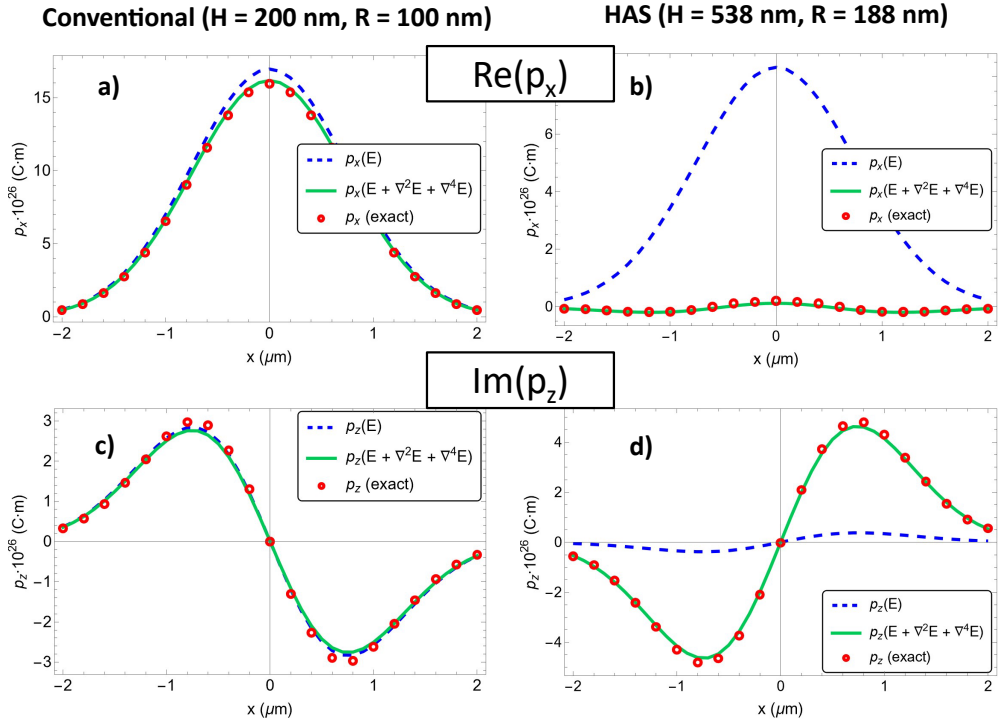

Figure S9. Electric dipole components $Re(p_x)$ and $Im(p_z)$ induced in the nanoparticle in the conventional scattering regime (a, c) and in the HAS (b,d). The blue dashed curves show the result of calculation using Eq. (2) (i.e., including response only to the incident field); the green solid curves show the result of calculation using nonlocal approach based on Eq. (21) up to the forth derivative; and the red markers give the full-wave result based on Table 1.

Figure S9 shows that in the conventional scattering regime (a), the px component is proportional to the incident electric field and has the shape of a Gaussian beam. In this case, the linear approximation (blue dashed line) is sufficient and gives a good match with full-wave simulation result (curve with red markers), while taking into account higher-order terms does not significantly change the result (green curve).

The same result is implied by the conventional approximation. For a homogenous isotropic particle dipole polarizability $\alpha_p$ is a diagonal tensor (see [19] for details). Thus, we can write: $p_x(x) = \alpha_p^{xx} E_x(x)$, where $E_x(x)$ is a distribution of an incidence electric field, and the polarizability factor $\alpha_p^{xx}$ is solely a function of frequency, but not the coordinates of the particle.

Unlike in the previous case, the situation in HAS is changing significantly (Figure S9, b). The shape of the $p_x$ component in full-wave simulation (curve with red markers) differs from the Gaussian profile so the classical approximation (blue dashed line) cannot describe it. To accurately match the full-wave calculation, it is necessary to take into account terms up to the 4th derivatives (green solid line).

### 4.5 Electric dipole force

After the examination of electric dipole in the hybrid anapole state we can explain the shape of the first term in formula (17), which reflects the contribution of electric dipole to the force. It consists of 3 components proportional to $p_x$, $p_y$ and $p_z$. Due to the symmetry of our system, $p_y = 0$. Figure S10 shows the contributions of $p_x$ and $p_z$ to the total electric dipole force in the conventional regime (nanocylinder with R = 100 nm, H = 200 nm) and in the hybrid anapole state (R = 188 nm, H = 538 nm):

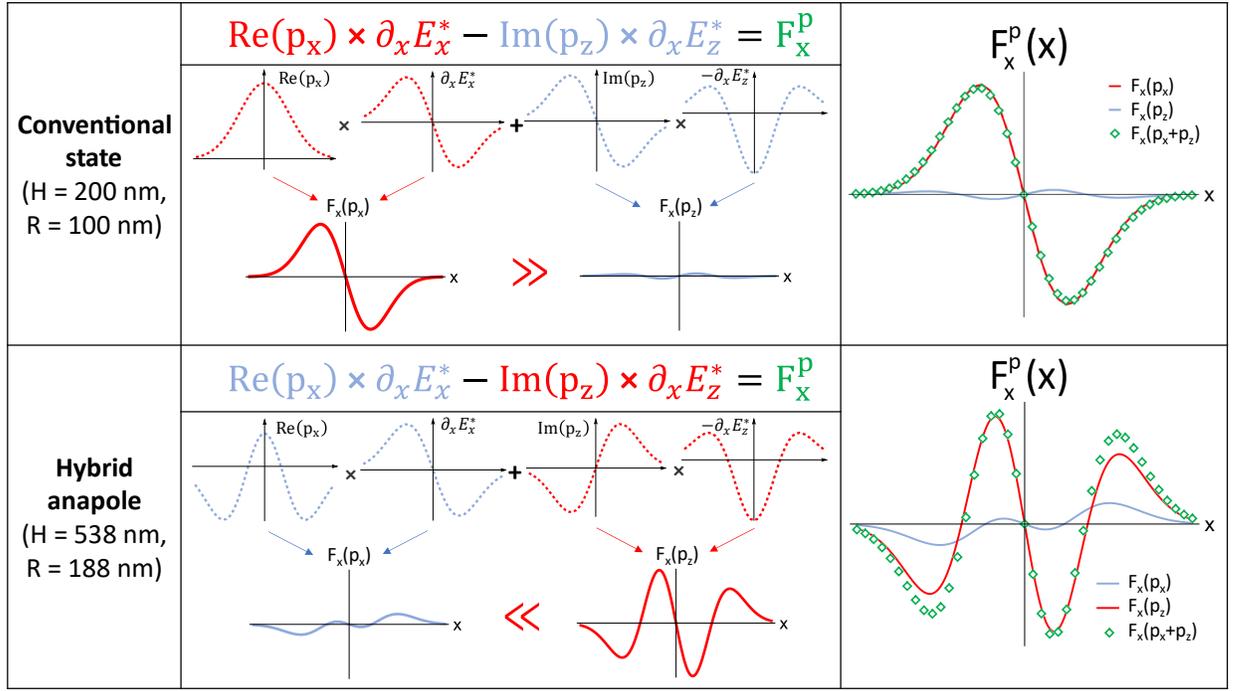

Figure S10. Electric dipole component of the transverse Fx force in the case of conventional trapping (upper panel) and HAS (lower panel). Central panel shows the components of the dipole multiplied by the field derivatives that contribute to the total force. The results are shown below them. Red curves depict the strongest components, light blue curves – the weakest. Right panel compares contributions from px, pz, and the total electric dipole force

Figure S10 reveals a pronounced difference in the contributions of dipole components to the total optical force in different scattering regimes. In the conventional case, the $p_x$ component dominates over $p_z$, while in the hybrid anapole state, the inverse is true. This example explains another mechanism of non-classical force formation in addition to the nonlocal response: within the HAS, the $p_x$ component decreases, while $p_z$ (which is proportional to the imagine component of Gaussian beam – $E_z$) remains significant, thereby becoming the dominant contributor that forms a force profile with 2 trapping points along the x-axis. In the conventional regime, the contribution of the $p_z$ to the force is negligible compared to $p_x$ so classical trapping is realized. The analysis of other multipoles allowed us to identify the components dominant in the conventional regime (e.g., $p_x, Q^e_{xz}, Q^m_{yz}$) and in the HAS (e.g., $p_z, Q^e_{xx}, Q^e_{zz}, Q^m_{zz}$).

*4.6 Interceptional components of force*

We now apply the aforementioned Taylor series expansion of the incident field to compute all interceptional components of the force in Eq. (17), both in the conventional scattering regime and in the HAS – Figure S11.

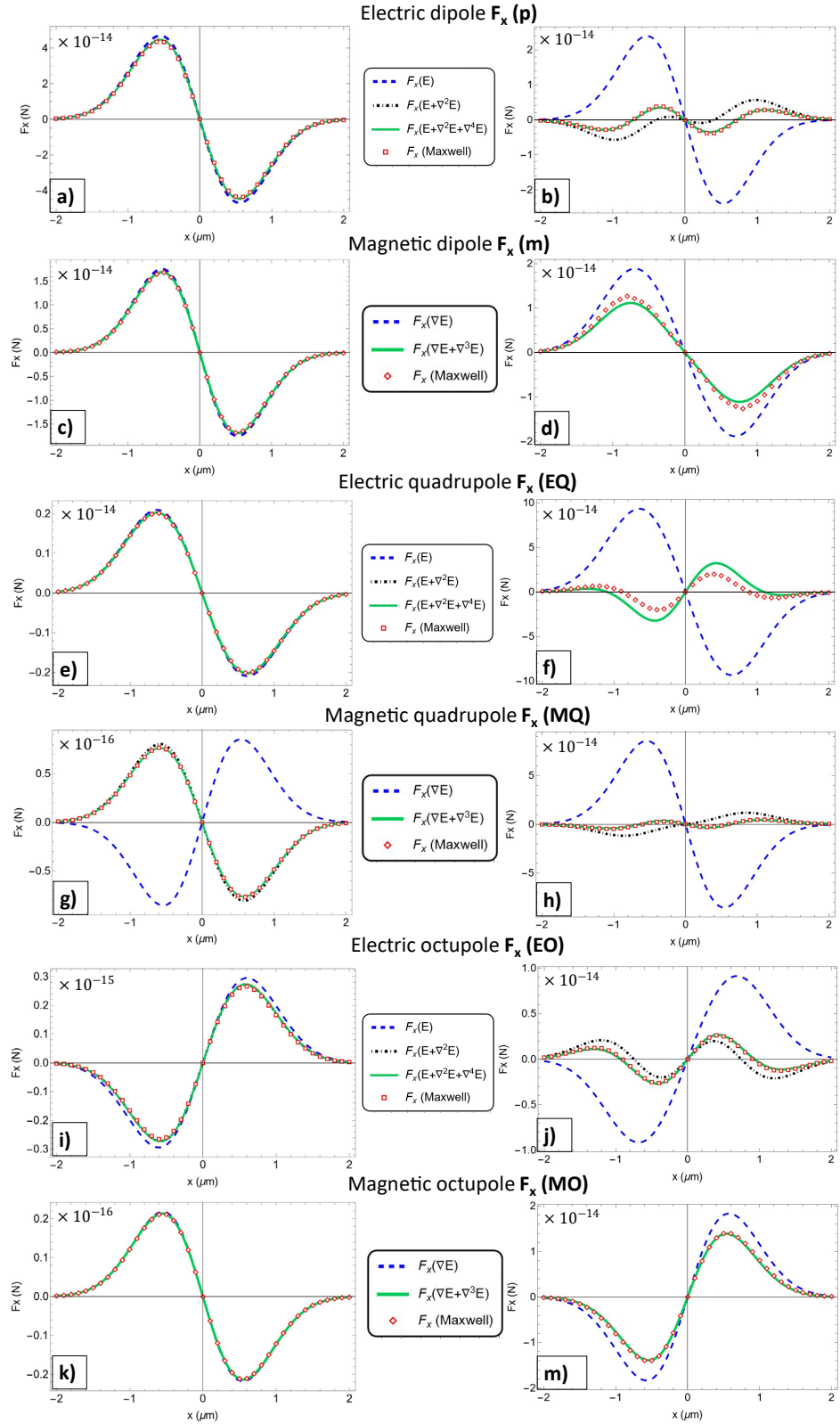

Figure S11. Interceptional components of the transverse $F_x$ force in the case of conventional trapping (panels a,c,e,g,i,k) and in the HAS (panels b,d,f,h,j,m). Blue dashed curves – the result of calculation using conventional approximation (i.e., including the response only to the incident field and its first derivatives); black dot-dashed curves – calculation using the Taylor series up to the second derivatives (for multipoles that respond to them), green solid curves – Taylor series up to the highest third or fourth derivatives for corresponding multipoles; and red markers give the full-wave result based on Table 1.

Figure S11 shows that in the conventional regime, it is sufficient to consider the response to the field and its first derivative to obtain a good match between the result and the forces calculated using full-wave simulation. The exception is the magnetic quadrupole, but its contribution is several orders of magnitude smaller than that of the other multipoles, so it is not noticeable when calculating the total force.

In the HAS, the situation is markedly different. To achieve agreement with full-wave simulation, it is necessary to consider the response to derivatives up to the 3$^{rd}$ order for "odd" multipoles (p, MQ, EO) and up to the 4$^{th}$ order for "even" multipoles (m, EQ, MO). This is a new optomechanical property that distinguishes hybrid anapole from conventional scattering regimes.